\documentclass[11pt]{article}
\usepackage{amsmath,amsthm,amssymb,cite}
\usepackage{psfrag}
\usepackage{graphicx}
\newcommand{\nc}{\newcommand}
\nc{\mref}[1]{(\ref{#1})}
\nc{\vt}{v_{2\gL_0}}
\nc{\vo}{v_{\gL_0}}
\nc{\vot}{v_{\gL_1+\gL_0}}
\nc{\vw}{v_{\gL_1}}
\nc{\ppmm}{\genfrac{}{}{-10pt}{10pt}{++}{--}}
\nc{\wom}[5]{\Omega\left(\left.\begin{array}{ll}{#1}&{#2}\\{#3}&{#4}\end{array}\right|{#5}\right)}
\nc{\com}[5]{\chi\left(\left.\begin{array}{ll}{#1}&{#2}\\{#3}&{#4}\end{array}\right|{#5}\right)}
\nc{\we}[5]{W\left(\left.\begin{array}{ll}{#1}&{#2}\\{#3}&{#4}\end{array}\right|{#5}\right)}
\nc{\ce}[6]{C^{#6}\left(\left.\begin{array}{ll}{#1}&{#2}\\{#3}&{#4}\end{array}\right|{#5}\right)}
\nc{\lmat}[6]{\ell_{#6}\left(\left.\begin{array}{ll}{#1}&{#2}\\{#3}&{#4}\end{array}\right|{#5}\right)}
\nc{\lmats}[5]{L\left(\left.\begin{array}{ll}{#1}&{#2}\\{#3}&{#4}\end{array}\right|{#5}\right)}
\nc{\hmat}[6]{h_{#6}\left(\left.\begin{array}{ll}{#1}&{#2}\\{#3}&{#4}\end{array}\right|{#5}\right)}
\nc{\hmats}[5]{H\left(\left.\begin{array}{ll}{#1}&{#2}\\{#3}&{#4}\end{array}\right|{#5}\right)}
\nc{\web}[5]{\overline{W}\left(\left.\begin{array}{ll}{#1}&{#2}\\{#3}&{#4}\end{array}\right|{#5}\right)}
\nc{\wep}[5]{W'\left(\left.\begin{array}{ll}{#1}&{#2}\\{#3}&{#4}\end{array}\right|{#5}\right)}
\nc{\wes}[5]{W^*\left(\left.\begin{array}{ll}{#1}&{#2}\\{#3}&{#4}\end{array}\right|{#5}\right)}
\nc{\wess}[5]{W^{**}\left(\left.\begin{array}{ll}{#1}&{#2}\\{#3}&{#4}\end{array}\right|{#5}\right)}
\nc{\cet}[7]{C^{#6}_{#7}\left(\left.\begin{array}{ll}{#1}&{#2}\\{#3}&{#4}\end{array}\right|{#5}\right)}
\nc{\bcet}[7]{\bar{C}^{#6}_{#7}\left(\left.\begin{array}{ll}{#1}&{#2}\\{#3}&{#4}\end{array}\right|{#5}\right)}
\nc{\wet}[7]{W^{#6}_{#7}\left(\left.\begin{array}{ll}{#1}&{#2}\\{#3}&{#4}\end{array}\right|{#5}\right)}
\nc{\bwet}[7]{\overline{W}^{#6}_{#7}\left(\left.\begin{array}{ll}{#1}&{#2}\\{#3}&{#4}\end{array}\right|{#5}\right)}
\nc{\wec}[7]{\widetilde{W}^{#6}_{#7}\left(\left.\begin{array}{ll}{#1}&{#2}\\{#3}&{#4}\end{array}\right|{#5}\right)}
\nc{\wgen}[6]{W^{#6}\left(\left.\begin{array}{ll}{#1}&{#2}\\{#3}&{#4}\end{array}\right|{#5}\right)}
\nc{\wgenp}[6]{W^{*{#6}}\left(\left.\begin{array}{ll}{#1}&{#2}\\{#3}&{#4}\end{array}\right|{#5}\right)}
\nc{\wo}[5]{\Omega\left(\left.\begin{array}{ll}{#1}&{#2}\\{#3}&{#4}\end{array}\right|{#5}\right)}
\nc{\wsgen}[8]{{#8}^{#6}_{#7}\left(\left.\begin{array}{ll}{#1}&{#2}\\{#3}&{#4}\end{array}\right|{#5}\right)}
\nc{\qbinom}[2]{{\genfrac{[}{]}{0pt}{}{{#1}}{{#2}}}_{q}}
\nc{\hhg}[4]{\phi\left({{{#1}\,\,\,{#2}}\atop{{#3}}};
                     {#4}\right)}
\nc{\fullhhg}[5]{ {_2}{\large \phi}_1 \left(   {   { {#1}\,\,\,{#2} } \atop { {#3} } };
                     {#4},{#5}\right)}
\nc{\bra}[1]{\langle #1 |}
\nc{\ket}[1]{| #1 \rangle}
\nc{\qp}[2]{({#1}\, ; \, {#2})_{\infty}}
\nc{\qpf}[1]{({#1}\, ; \, q^4)_{\infty}}
\nc{\pp}[1]{({#1}\, ; \, p)_{\infty}}
\nc{\qpp}[1]{({#1}\, ; \, p, q^4)_{\infty}}
\nc{\sect}{\section}
\nc{\ssect}{\subsection}
\nc{\sssect}{\subsubsection}
\nc{\ud}[1]{\underline{{#1}}}
\nc{\myra}[1]{\buildrel{{#1}}\over \longrightarrow}
\nc{\isomo}{\buildrel {\sim} \over \longrightarrow}
\nc{\Aff}{\operatorname{Aff}}
\nc{\ot}{\otimes}
\nc{\er}{\end{array}}
\nc{\bev}[1]{\begin{equation}\begin{array}{#1}}
\nc{\eeq}{\end{equation}}
\nc{\be}{\begin{eqnarray}}
\nc{\ee}{\end{eqnarray}}
\nc{\ben}{\begin{eqnarray*}}
\nc{\een}{\end{eqnarray*}}
\nc{\bec}{\begin{equation}\begin{array}{lll}}
\nc{\eec}{\end{array}\end{equation}}
\nc{\ed}{\end{document}}
\nc{\half}{\ensuremath{\frac{1}{2}}}
\nc{\Hom}{\operatorname{Hom}}
\nc{\End}{\operatorname{End}}
\nc{\vac}{|\textrm{vac}\rangle}
\nc{\tvac}{|\widetilde{\textrm{vac}}\rangle}
\nc{\dvac}{\langle\textrm{vac}}
\nc{\dtvac}{\langle\widetilde{\textrm{vac}}}
\nc{\id}{\mathbb{I}}
\nc{\ra}{\rightarrow}  
\nc{\lra}{\longrightarrow}
\nc{\uqp}{U^{\prime}_q (\widehat{sl}_2)}
\nc{\uqbp}{U_q (b_+)}
\nc{\uqbm}{U_q (b_-)}
\nc{\ub}{U^{\prime}_q (b_+)}
\nc{\vsl}{V(\sigma(\lambda))}
\nc{\vl}{V(\lambda)}  
\nc{\bu}{\bullet}
\nc{\an}{{\ell}}
\nc{\slth}{\widehat{\mathfrak{sl}}_2\hskip 1pt}
\newcommand{\uq}{U_q\bigl(\slth\bigr)}
\nc{\ws}{\;\;}
\nc{\qu}{{1\ov 4}}
\nc{\hif}{\hb{ if }}
\nc{\hev}{\hb{ is even }}
\nc{\hod}{\hb{ is odd }}
\nc{\Tr}{{\rm Tr}}
\nc{\ad}{{\rm Ad}}
\nc{\hb}{\hbox}
\nc{\nn}{\nonumber} 
\nc{\curlra}{\buildrel{\sim}\over\longrightarrow}
\nc{\epp}{\varepsilon^{\prime}} 
\nc{\ol}{\overline}
\nc{\pl}{\prod\limits} 
\nc{\sli}{\sum\limits} 
\nc{\nin}{\noindent}

\nc{\ga}{\alpha}
\nc{\gb}{\beta}
\nc{\gd}{\delta}
\nc{\gep}{\varepsilon}
\nc{\gz}{\zeta}
\nc{\gt}{\theta}
\nc{\gk}{\kappa}
\nc{\gl}{\lambda}
\nc{\gp}{\phi}
\nc{\gs}{\sigma}
\nc{\go}{\omega}
\nc{\gn}{\nu}
\nc{\gr}{\rho}

\nc{\gou}{\underline{\go}}
\nc{\un}{\underline{n}}
\nc{\um}{\underline{m}}
\nc{\uw}{\underline{w}}

\nc{\s}{\sigma}
\nc{\ep}{\varepsilon}
\nc{\z}{\zeta}
\nc{\g}{\gamma}
\nc{\zi}{\zeta^{-1}}

\nc{\gG}{\Gamma}
\nc{\gD}{\Delta}
\nc{\gT}{\Theta}
\nc{\gL}{\Lambda}
\nc{\gO}{\Omega}
\nc{\gP}{\Phi}

\nc{\cL}{\mathcal{L}}
\nc{\cF}{\mathcal{F}}
\nc{\cP}{\mathcal{P}}
\nc{\cS}{\mathcal{S}}
\nc{\cN}{\mathcal{N}}
\nc{\cD}{\mathcal{D}}
\nc{\cH}{\mathcal{H}}
\nc{\cO}{\mathcal{O}}
\nc{\cT}{\mathcal{T}}
\nc{\cQ}{\mathcal{Q}}
\nc{\cW}{\mathcal{W}}
\nc{\cR}{\mathcal{R}}

\nc{\C}{\mathbb{C}}
\nc{\Q}{\mathbb{Q}}
\nc{\R}{\mathbb{R}}
\nc{\Z}{\mathbb{Z}}
\nc{\N}{\mathbb{N}}

\nc{\fg}{\mathfrak{g}}

\nc{\barx}{\bar{x}}
\nc{\bi}{\bar{i}}
\nc{\bj}{\bar{j}}
\nc{\bgr}{\bar{\rho}}
\nc{\bA}{\bar{\alpha}}
\nc{\bB}{\bar{\beta}}
\nc{\bC}{\bar{\gamma}}
\nc{\by}{\bar{y}}
\nc{\brv}{\overline{V}}
\nc{\brp}{\overline{P}}

\nc{\tf}{\tilde{f}}
\nc{\te}{\tilde{e}}
\nc{\ts}{\tilde{s}}
\nc{\tgP}{\widetilde{\Phi}}
\nc{\tgPs}{\tilde{\Psi}}
\nc{\tgn}{\tilde{\nu}}
\nc{\tgl}{\tilde{\lambda}}
\nc{\tge}{\tilde{\eta}}
\nc{\txi}{\tilde{\xi}}
\nc{\tep}{\tilde{\epsilon}}

\nc{\cB}{\check{b}}
\nc{\cOm}{\check{\Omega}}

\nc{\goto}{\mapsto}
\nc{\embed}{\hookrightarrow}
\nc{\rien}{\emptyset}
\nc{\lb}[1]{\label{#1}}
\nc{\Nt}{\frac{N}{2}}
\nc{\vn}{\hspace*{-33truemm}}
\nc{\vm}{\hspace*{-0truemm}}
\nc{\ti}{t^{-1}}
\nc{\vb}{v^{(1)}}
\nc{\vbn}{v^{(n)}}
\nc{\ur}{\underline{r}}
\nc{\us}{\underline{s}}
\nc{\up}{\underline{p}}
\nc{\bp}{\bar{p}}
\nc{\bpi}{\bar{p}^{(i)}}
\nc{\bpip}{\bar{p}^{(i+1)}}
\nc{\vz}{V^{(1)}_z}
\nc{\vzn}{V^{(n)}_z}
\nc{\vzo}{V^{(1)}_1}
\nc{\piz}{\pi_z^{(1)}}
\nc{\pizn}{\pi_z^{(n)}}
\nc{\pis}{\pi_{(z,\us)}}
\nc{\bW}{\overline{W}}
\nc{\bQ}{\overline{Q}}
\nc{\tQ}{\widetilde{Q}}
\nc{\bT}{\overline{T}}
\nc{\note}[1]{\vspace*{-5mm}\marginpar[left]{\scriptsize\bf{#1}}}
\nc{\mynote}[1]{\nin{\bf Note:{#1}}\newline}
\nc{\eqdef}{:=}
\nc{\lu}{^{(\lambda)}}
\nc{\vone}{v_{\Lambda_1}}
\nc{\vzero}{v_{\Lambda_0}}

\def\bea{\begin{eqnarray}}
\def\eea{\end{eqnarray}}

\def\ep{\varepsilon}

\newtheorem{theorem}{Theorem}[section]

\newtheorem{example}[theorem]{Example}
\def\by{{\mbox{\boldmath $y$}}}

\def\bpm{\begin{pmatrix}}
\def\epm{\end{pmatrix}}
\def\bdet{\left|\begin{array}}
\def\edet{\end{array}\right|}

\nc{\dx}[1]{\frac{d{#1}}{dx}}
\nc{\ddx}[1]{\frac{d^2{#1}}{dx^2}}
\nc{\dt}[1]{\frac{d{#1}}{dt}}
\nc{\ddt}[1]{\frac{d^2{#1}}{dt^2}}

\newcounter{quest}
\newlength{\questoffset}
\nc{\pd}{\partial}
\nc{\nl}{\newline}
\nc{\vp}{\vspace*{3mm}}
\nc{\soln}{\noindent {\it Solution}:\,}
\nc{\pro}{\noindent {\it Proof}\,:\,}
\nc{\bex}{\begin{example}}
\nc{\eex}{\end{example}}

\nc{\mitem}{\noindent}

\newcommand{\ena}{\end{eqnarray}}

\def\cip(#1){(#1;p,q^4)_\infty}
\def\z{\zeta}

\def\Lpm#1{\mathrel{\mathop{\kern0pt L^\pm}\limits^#1}}
\def\L#1{\mathrel{\mathop{\kern0pt L}\limits^#1}}
\def\Lm#1{\mathrel{\mathop{\kern0pt L^-}\limits^#1}}
\def\Lp#1{\mathrel{\mathop{\kern0pt L^+}\limits^#1}}

\def\H{{\cal H}}

\begin{document}
\bibliographystyle{unsrt}
\begin{flushright}
\end{flushright}
\begin{center}
{\LARGE \bf Correlation Functions and the Boundary qKZ Equation in a Fractured XXZ Chain}\\[10mm]
{\Large \bf 
Robert Weston}\\[3mm]
{\it Department of Mathematics, Heriot-Watt University,\\
Edinburgh EH14 4AS, UK.
}\\[5mm]
October 2011\\[10mm]

\end{center}
\begin{abstract}
\noindent 
We consider correlation functions of the form 
$\langle \hb{vac}| {\cal O} |\hb{vac}\rangle'$, where $|\hb{vac}\rangle$ is the vacuum eigenstate of an infinite antiferromagnetic XXZ chain, $|\hb{vac}\rangle'$ is the vacuum eigenstate of an infinite XXZ chain which is split in two, and ${\cal O}$ is a local operator. The Hamiltonian of the split chain has no coupling between sites $1$ and $0$ and has a staggered magnetic field at these two sites; it  arises from a tensor product of left and right boundary transfer matrices. We find a simple, exact expression for $\langle \hb{vac}|\hb{vac}\rangle'$ and an exact integral expression for general $\langle \hb{vac}| {\cal O} |\hb{vac}\rangle'$  using the vertex operator approach. We compute the integral when ${\cal O}=\sigma^z_1$ and find a conjectural expression that is analogous to the known formula for the XXZ spontaneous magnetisation and reduces to it when the magnetic field is zero. We show that correlation functions obey a boundary qKZ equation of a different level to the infinite
XXZ chain with one boundary. 
\end{abstract}
\nopagebreak

\section{Introduction}
\setcounter{equation}{0}
The periodic XXZ chain is a quantum system with Hamiltonian 
\bea H_{XXZ}=-\half \sli_{j=1}^{N} \left (\sigma_{j+1}^x \sigma_{j}^x + \sigma_{j+1}^y \sigma_{j}^y+\Delta \sigma_{j+1}^z \sigma_{j}^z\right),\lb{xxz1}\eea
where $\sigma_j^\alpha$ denotes the Pauli matrix $\sigma^\alpha$ acting at the $j$'th site of the chain. We label our sites from right to left and 
identify the $N+1$'th site with the 1st site of the chain.
This simple model has been the subject of a huge amount of attention from the time of Bethe \cite{Bethe31} until the present day 
\cite{MR2583103}. There are two main reasons for this interest. Firstly, the model is quantum integrable and 
has proved to be both a source and test-bed for a set of theoretical techniques applicable to such systems. These approaches include the Bethe Ansatz, quantum inverse scattering \cite{Korepin}, the corner transfer matrix and Q operator \cite{Bax82}, quantum field theory \cite{Aff89}, quantum groups \cite{Jim85,Dri85} and the vertex operator approach \cite{Daval,JM}. Secondly, the XXZ model is of physical interest; it was developed as an idealised model of ferromagnetic systems and has recently found  direct experimental realisation in a range of quasi-one-dimensional materials \cite{Goff95,Nagler83,PhysRevLett.91.090402}. The Hamiltonian has also made a guest appearance in ${\cal N}=4$ super Yang Mills theory in the last decade \cite{Beisertreview}.

One of the key goals of all the theoretical work on the XXZ chain has been to exploit its quantum integrability in order to compute exact expressions for correlation functions. This goal was finally achieved for the infinite lattice in the 1990s. Two independent approaches were developed. The first was invented by the Kyoto group and deals with the infinite system with antiferromagnetic boundary conditions \cite{Daval,JM}. In this case, the Hamiltonian commutes with the action of the quantum affine algebra $\uq$. This observation ultimately allowed the Kyoto group to produce an entirely representation theoretic expression for arbitrary correlation functions. We will refer to the techniques of the Kyoto group as the `vertex operator approach'. The second approach, developed by the Lyon group, involved a direct Bethe-Ansatz computation of correlation functions for the finite-size system and a careful thermodynamic limit \cite{MR1702631,MR1741654}.

In this paper, we study the XXZ model in a setting that we refer to as a {\it fractured} XXZ chain. We consider the eigenstates of two separate Hamiltonia. The first Hamiltonian is simply the infinite-size antiferromagnetic Hamiltonian
\ben H=-\half \sli_{j\in \Z} \left (\sigma_{j+1}^x \sigma_{j}^x + \sigma_{j+1}^y \sigma_{j}^y+\Delta 
\sigma_{j+1}^z \sigma_{j}^z\right),\quad
\Delta=\frac{q+q^{-1}}{2}, \ws -1<q<0,\een
that acts on the space with antiferromagnetic boundary conditions $+-+-+-$ or $-+-+-+$ at plus and minus infinity (we use $+$ and $-$ to refer to two eigenstates of $\sigma^z$ -  see Section 2 for a more precise definition of this space). 
This Hamiltonian is the one already dealt with by the Kyoto group and we simply recall the results that we need.  
In particular, 
labelling the two antiferromagnetic boundary conditions by $i=0$ and $i=1$, we make use of the two corresponding lowest energy eigenstates $\vac_{(i)}$ constructed in \cite{JM}. The second Hamiltonian we consider, which acts on the same space as above, is a split Hamiltonian 
\ben H'=H_L+H_R,\een where the left and right Hamiltonia are
\ben
H_L&=& -\half \sli_{j\geq 1} \left (\sigma_{j+1}^x \sigma_{j}^x + \sigma_{j+1}^y \sigma_{j}^y+\Delta \sigma_{j+1}^z \sigma_{j}^z\right)+h \sigma_1^z,\\
H_R&=& -\half \sli_{j\leq 0} \left (\sigma_{j}^x \sigma_{j-1}^x + \sigma_{j}^y \sigma_{j-1}^y+\Delta \sigma_{j}^z \sigma_{j-1}^z\right)-h \sigma_0^z.\een
With this Hamiltonian, there is split or fracture between the sites at position 1 and 0, with no nearest-neighbour interaction between them. There is also
a staggered magnetic field $h$ that acts at the boundaries of the fracture. Let us label the vacuum of $H'$ by $\vac'_{(i)}$. This state will of course be simply a tensor product of the respective lowest eigenstates of $H_L$ and $H_R$. 

The key results of this paper are exact expressions for \bea {_{(i)}}\!\langle\hb{vac}\vac'_{(i)},\lb{fidintro}\eea and for all correlation functions of the form
\bea {_{(i)}}\!\langle\hb{vac}| {\cal O}\vac'_{(i)},\lb{cfnintro}\eea
where ${\cal O}$ is any local operator. 
We obtain these expression by developing the vertex operator approach. 

There are several reasons why we look at the matrix elements \mref{fidintro} and \mref{cfnintro} for this system. Firstly, they are `natural' objects to consider in the language of the related 6-vertex model. We shall develop this point of view throughout the paper, 
but in brief, the matrix element  
${_{(i)}}\!\langle\hb{vac}\vac'_{(i)}$ 
is related to the partition function of the infinite lattice shown in Figure \ref{fracpfn}. As such, it can be expressed as a matrix element
of two boundary, and two bulk corner transfer matrices representing the four quadrants of the lattice. Corner transfer matrices of both boundary and bulk types have  been considered previously in the vertex operator approach \cite{JKKMW,JM}, and it is very natural to put the two types together to form the fractured 
partition function of Figure  \ref{fracpfn}.

The second motivation for our interest in these fractured systems is their potential physical significance. One application is in the description of a {\it local quantum quench} in which a spin chain that consists of two separated domains for time $t<0$ is joined at time $t=0$.
Such local quenches have been investigated using conformal field theory techniques in \cite{1742-5468-2007-06-P06005,CC07,1742-5468-2011-03-L03002,DubStep11}.
In such a  context, the squared modulus of the overlap $\langle \hb{vac}|\hb{vac}\rangle'$, which we compute exactly, is referred to as the {\it fidelity} and plays a role in understanding the dynamical behaviour of quantum entanglement \cite{PhysRevLett.106.150404,DubStep11}. 

A final reason for our interest in this fractured geometry is the potential connection with existing work on two-dimensional models on a wedge of angle $\alpha$. Both massless and massive systems have been considered in such wedge geometries in \cite{0305-4470-16-15-026,0305-4470-17-17-005} and \cite{MR775790}, and there exists the possibility of establishing a relation with our work when $\alpha\ra 2\pi$.

In this paper we construct the matrix element ${_{(i)}}\langle\hb{vac}| {\cal O}\vac'_{(i)}$ as a specialisation of a function $P^{(i)}(\z_1,\z_2,\cdots,\z_N)$ defined as a matrix element of a product of vertex operators. We find  that $P^{(i)}(\z_1,\z_2,\cdots,\z_N)$ satisfies a boundary qKZ equation similar to the form found in \cite{JKKMW}, but at a different level. We use a free-field realisation to derive a general multiple-integral expression for  $P^{(i)}(\z_1,\z_2,\cdots,\z_N)$. Specialising to the case of the magnetisation, 
corresponding to the choice $\cO=\sigma_1^z$, we integrate the resulting expression to produce a q-expansion to order $q^{96}$. Based on this expansion we make the following conjecture for the exact expression
\bea -\,{_{(0)}}\!\langle\hb{vac}| \sigma_1^z\vac'_{(0)}= 1+2(1-r) \sli_{n=1}^{\infty} \frac{(-q^2)^n}{(1-rq^{4n})},
\label{mag}\eea
where $-1\leq r\leq 1$ is related to the magnetic field $h\geq 0$ by
\ben  h= \frac{(q^2-1)}{4q} \frac{1+r}{1-r}.\een
Note that $r=-1$ corresponds to the magnetic field $h=0$ and $r=1$ corresponds to $h=\infty$. Thus we have the special cases
\ben -\, {_{(0)}}\!\langle\hb{vac}| \sigma_1^z\vac'_{(0)}=\begin{cases}
 1+4 \sli_{n=1}^{\infty} \frac{(-q^2)^n}{(1+q^{4n})}=\frac{(q^2;q^2)_\infty^2}{(-q^2;q^2)_\infty^2},\quad &h=0,\\
1,\quad &h=\infty\end{cases}.\een
The first expression coincides with the known bulk spontaneous magnetisation \cite{Bax82}. We argue that this result and the $h=\infty$ limit are as physically expected. We also make some observations about the special nature of the point $r=0$.

The contents of the paper are as follows: in Section 2, we provide the definition of the model. In Section 3, we describe the realisation of the model within the vertex operator approach. We construct the vacuum and excited eigenstates for the two
 Hamiltonia $H$ and $H'$ and give a vertex operator expression for correlation functions. We then demonstrate how these same correlation function expressions arise from a corner transfer matrix approach. We go on to show that the correlation functions obey a version of the boundary qKZ equation.
In Section 4, we make use of the free-field realisation to produce a general integral formula for correlation functions and show how its specialisation leads to the result \mref{mag}. Finally, we summarise and make some concluding comments in Section 5.

\section{The Model}
\setcounter{equation}{0}
In this section, we define our fractured XXZ/6-vertex model. All conventions closely follow those of the previous work on the vertex operator approach to the bulk \cite{Daval,JM} and boundary \cite{JKKKM,JKKMW} XXZ models. Our starting point is to consider a vertex model defined in terms of the bulk and boundary Boltzmann weights $R(\z)$ and $K(\z;r)$ defined in \cite{JKKKM} and reproduced in Appendix A. For convenience, we also define $K_\bullet(\zeta)=K(\zeta;r)$  and $K_\circ(\zeta)=K(-q^{-1}\zeta^{-1};r)$. Let $V$ be the two-dimensional space $V=\C v_+\oplus \C v_-$. The finite-lattice bulk and boundary transfer matrices are then operators $V^{\ot N}\ra V^{\ot N}$ defined by
\bea T^{fin}(\z)&=&\Tr_{V_0}\left( {\cal T}(\z)\right),\nn\\ \hb{and}\quad 
T_B^{fin}(\zeta)&=& \hb{Tr}_{V_0}(K_\circ(\zeta) {\cal T}^{-1}(\zeta^{-1}) K_\bullet (\zeta) {\cal T}(\zeta)),\lb{ptm}\\
\hb{where}\quad  {\cal T}(\zeta)&=& R_{01}(\z) R_{02}(\z)\cdots R_{0N}(\z)\in \hb{End}(V_0\ot V_N \ot \cdots\ot V_1).\nn\eea 
The boundary transfer matrix $T_B^{fin}(\zeta)$ \cite{SKl87} is represented by Figure 1, in which we make use of the pictorial conventions of Appendix A. We view this and other transfer matrices in this paper as acting in a downwards direction.
\begin{figure}[htbp]\label{btrans}
\centering
\includegraphics[width=4cm]{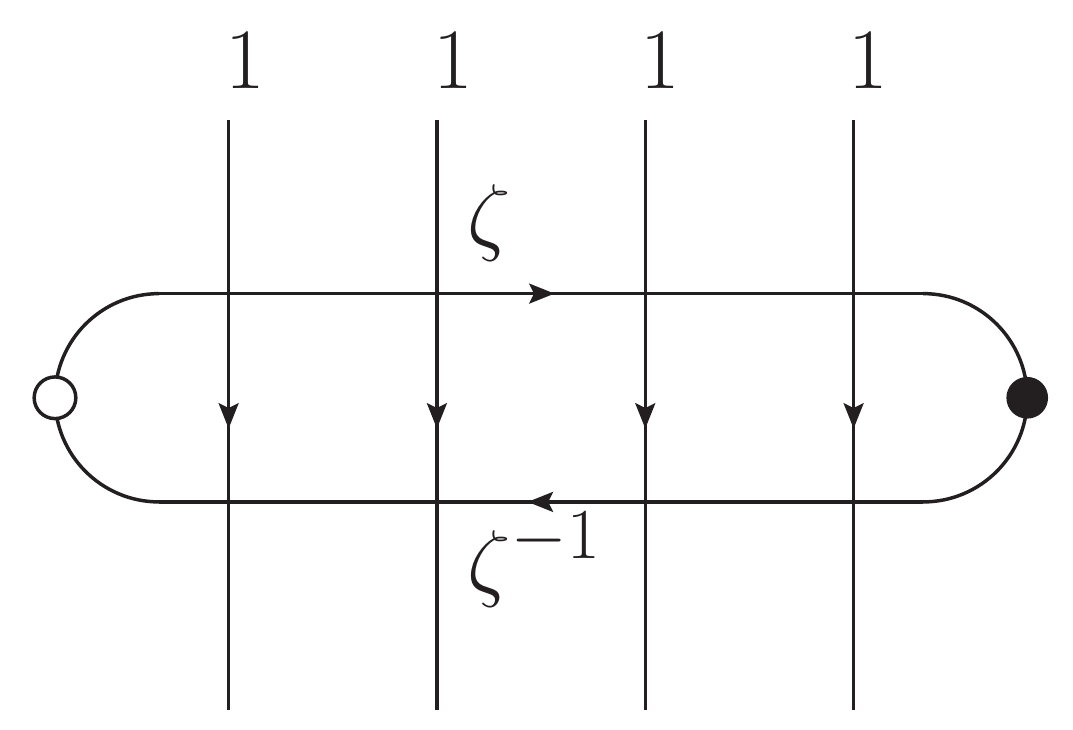}
\caption{The Transfer Matrix $T_B^{fin}(\z)$}
\label{frm1}
\end{figure}

The two transfer matrices are related to bulk and boundary Hamiltonia $H^{fin}$ and $H_B^{fin}$ by
\bea 
H^{fin}&:=&-\half \sli_{j=1}^{N} \left (\sigma_{j+1}^x \sigma_{j}^x + \sigma_{j+1}^y \sigma_{j}^y+\Delta \sigma_{j+1}^z \sigma_{j}^z\right)\\&=&
\frac{1-q^2}{2q} \frac{d}{d\z} \log T^{fin}(\z)|_{\z=1}+constant,\nn\\[2mm]
H_B^{fin}&:=&
-\half \sli_{j=1}^{N-1} \left (\sigma_{j+1}^x \sigma_{j}^x + \sigma_{j+1}^y \sigma_{j}^y+\Delta \sigma_{j+1}^z \sigma_{j}^z\right)+h \sigma_1^z -h \sigma_N^z\lb{hfinbound}\\
 &=&
\frac{1-q^2}{4q} \frac{d}{d\z} T_B^{fin}(\z)|_{\z=1}+constant,\nn\\[2mm]
\hb{where}\ws h&=& \frac{(q^2-1)}{4q} \frac{1+r}{1-r}.\lb{magfield}\eea

The  partition function of interest to us is the infinite-lattice limit of the one shown in Figure 2.
\begin{figure}[htbp]
\centering
\includegraphics[width=7cm]{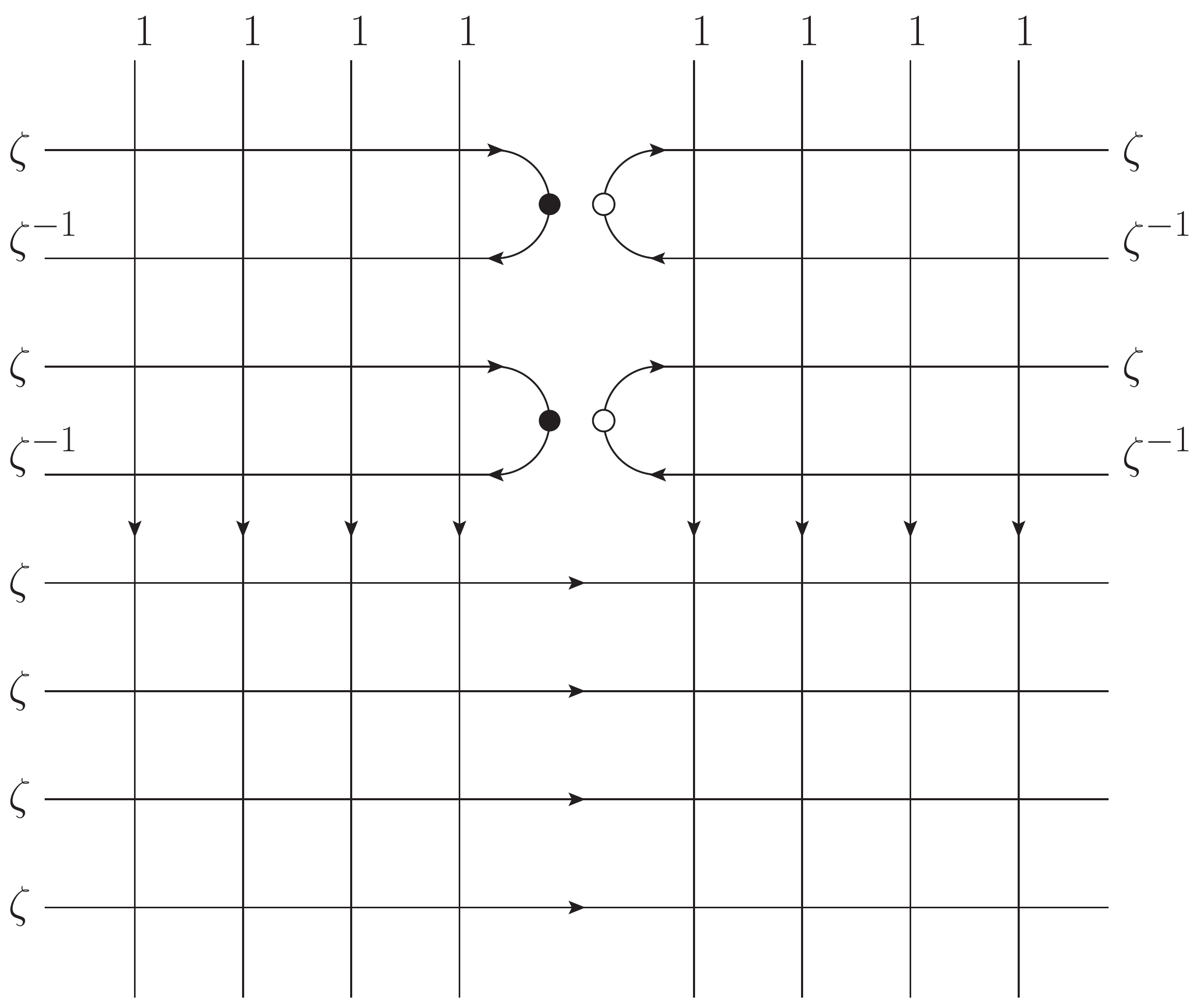}
\caption{The Partition Function of the Fractured Model}
\label{fracpfn}
\end{figure}
We specify two different boundary conditions which we label by $i=0$ and $i=1$. For $i=0$, the partition function is the sum over spin configurations of Figure \ref{fracpfn} that are fixed at finite but arbitrarily large distances from the centre of the lattice to the ground state shown in Figure \ref{fracgs}. For $i=1$, the configurations are restricted by the corresponding ground-state with $+\leftrightarrow -$.
\begin{figure}[htbp]
\centering
\includegraphics[width=7cm]{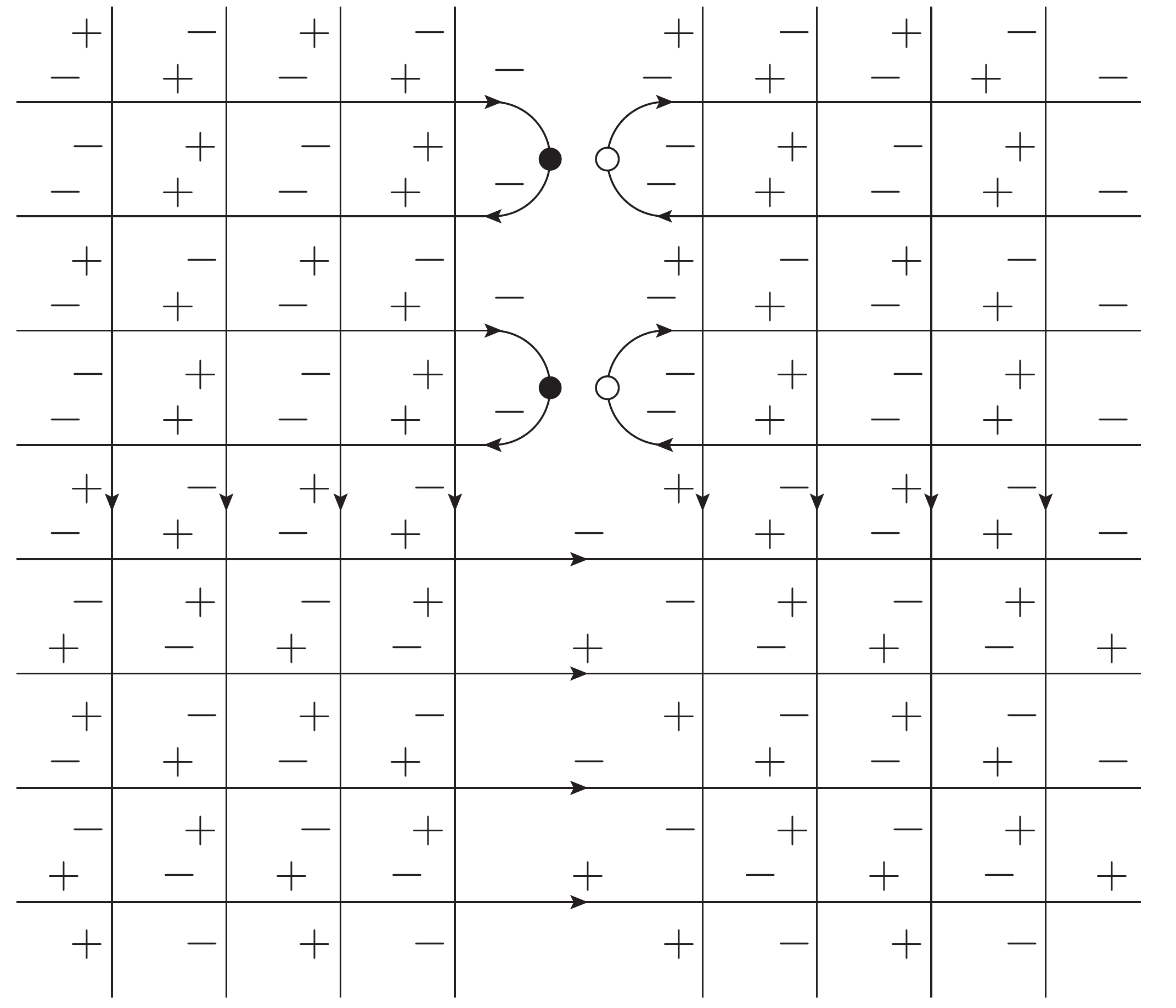}
\caption{The $i=0$ Ground State}
\label{fracgs}
\end{figure}

\begin{figure}[htbp]
\centering
\includegraphics[width=7cm]{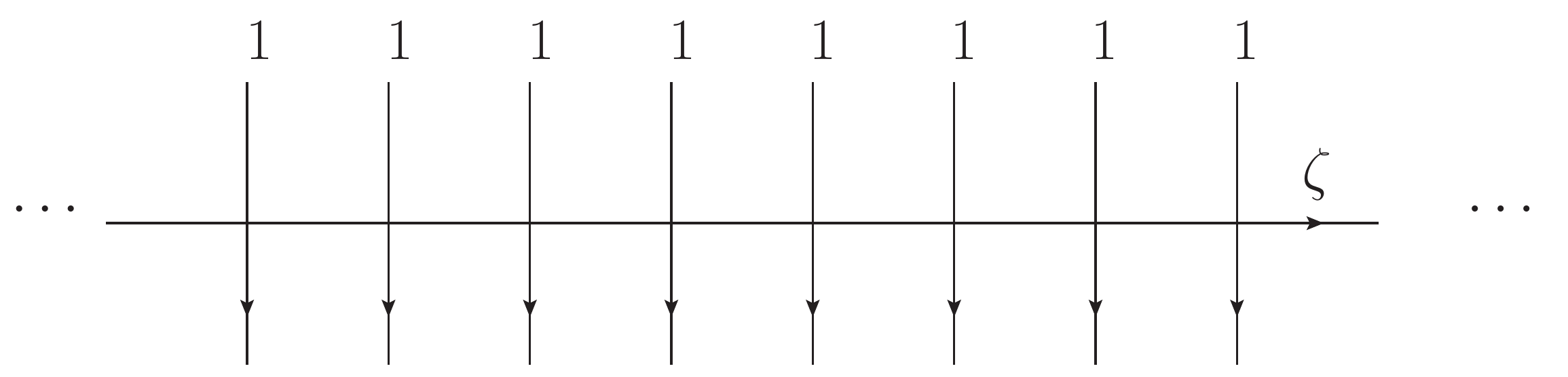}
\caption{The Bulk Transfer Matrix $T(\z)$}
\label{bulktrans}
\end{figure}

\begin{figure}[htbp]
\centering
\includegraphics[width=7cm]{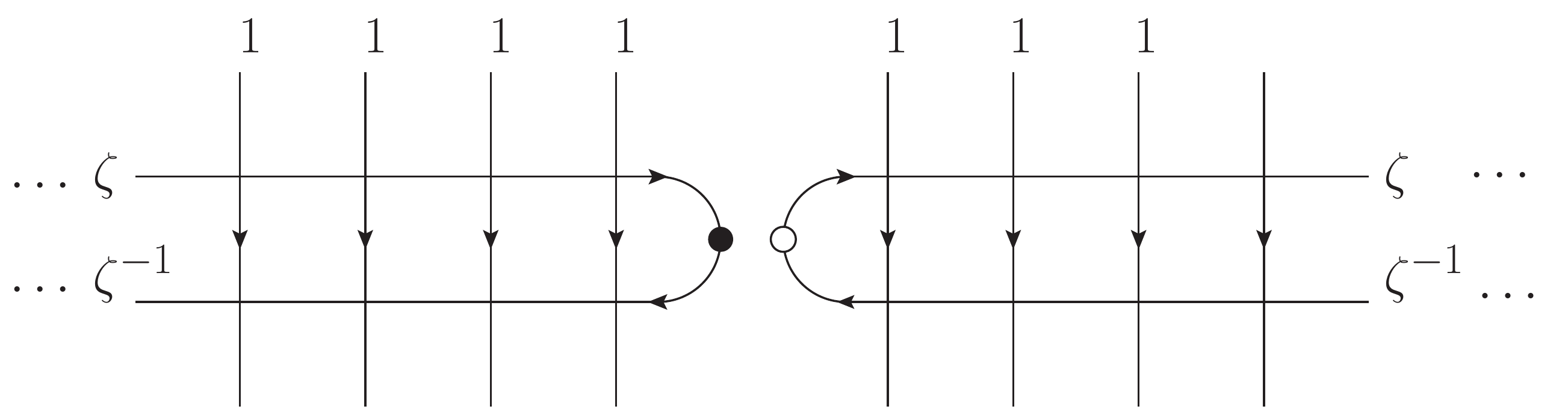}
\caption{The Fracture Transfer Matrix $T'(\z)=T_L(\z)\ot T_R(\z)$}
\label{fractrans}
\end{figure}

A key feature of the vertex operator approach is that it formulates such partition functions in terms of operators that act on the half-infinite left and right vector spaces
$\cH_L^{(i)}$ and $\cH_R^{(i)}$ defined by
\ben \cH_L^{(i)}&=&\hb{Span}\{\cdots \ot v_{\ep(2)}\ot v_{\ep(1)}| \ep(n)=(-1)^{n+i}, n\gg 0\},\\
\cH_R^{(i)}&=&\hb{Span}\{ v_{\ep(0)}\ot v_{\ep(-1)}\ot\cdots | \ep(n)=(-1)^{n+i}, n\ll 0 \}.\een
If we identify a dual vector  $v^*_{\pm}$ with $v_{\mp}$, then we can identify $\cH_R^{(i)}$ with $\cH_L^{*(i)}$. The full horizontal space associated with the partition function of Figure 2 is
\bea\cF^{(i)}=\cH_L^{(i)}\ot \cH_R^{(i)}\simeq \cH_L^{(i)}\ot \cH_L^{*(i)}\simeq \End(\cH_L^{(i)}).
\label{vspaceident}\eea
There are then two different horizontal transfer matrices associated with the partition function represented by Figure 2: the bulk transfer matrix $T(\zeta):\cF^{(i)}\ra \cF^{(1-i)}$ of Figure \ref{bulktrans} associated with the lower half of Figure 2; and the transfer matrix $T'(\zeta):\cF^{(i)}\ra \cF^{(i)}$ of Figure \ref{fractrans} associated with the upper half. We can clearly write $T'(\z)$ as the tensor product 
$T'(\z)=T_L(\z)\ot T_R(\z)$ of the left and right transfer matrices $T_{L,R}(\z):\cH^{(i)}_{L,R}\ra \cH^{(i)}_{L,R}$ shown in Figure \ref{fractrans}.

Differentiating these transfer matrices, using the explicit form of $R(\z)$, $K_\circ(\z)$ and $K_\bullet(\z)$ given in Appendix A, 
gives the associated infinite-lattice Hamiltonia $H, H':\cF^{(i)}\ra \cF^{(i)}$ as follows:
\bea
H&:=& -\half \sli_{j\in\Z} \left (\sigma_{j+1}^x \sigma_{j}^x + \sigma_{j+1}^y \sigma_{j}^y+\Delta \sigma_{j+1}^z \sigma_{j}^z\right)\lb{bulkham}\\
&=& \frac{1-q^2}{2q} \frac{d}{d\z} \log T(\z)|_{\z=1}+constant,\\[2mm]
\hb{and   }H'&:=&H_L+H_R,\quad H_{L,R}:\H_{L,R}^{(i)}\ra \H_{L,R}^{(i)},\nn\\
&&\hspace*{-10mm}\hb{with} \quad H_L:= -\half \sli_{j\geq 1} \left (\sigma_{j+1}^x \sigma_{j}^x + \sigma_{j+1}^y \sigma_{j}^y+\Delta \sigma_{j+1}^z \sigma_{j}^z\right)+h \sigma_1^z,\lb{leftham}\\
&&\hspace*{-10mm}\hb{and} \quad H_R= -\half \sli_{j\leq 0} \left (\sigma_{j}^x \sigma_{j-1}^x + \sigma_{j}^y \sigma_{j-1}^y+\Delta \sigma_{j}^z \sigma_{j-1}^z\right)-h \sigma_0^z,\lb{rightham}\\
\hb{where }H'&=& \frac{1-q^2}{4q} \frac{d}{d\z} T'(\z)|_{\z=1}+constant.
\eea

The first task in this paper is to diagonalise the Hamiltonia $H$ and $H'$. Fortunately, 
this task has already been carried for both $H$ and $H_L$ by using the vertex operator approach to the bulk \cite{JM} and boundary \cite{JKKKM} XXZ chains. It is then a simple matter to also diagonalise $H_R$ by relating it to $H_L$; we describe the eigenstates in the next section. Denoting the unique, lowest-energy eigenstates of $H$ and $H'$ in the space $\cF^{(i)}$ by $\vac_{(i)}$ and $\vac'_{(i)}$, we can identify the value of the matrix element $_{(i)}\langle \hb{vac}\vac'_{(i)}$ with the partition function of Figure 2. 

\begin{figure}[htbp]
\centering
\includegraphics[width=7cm]{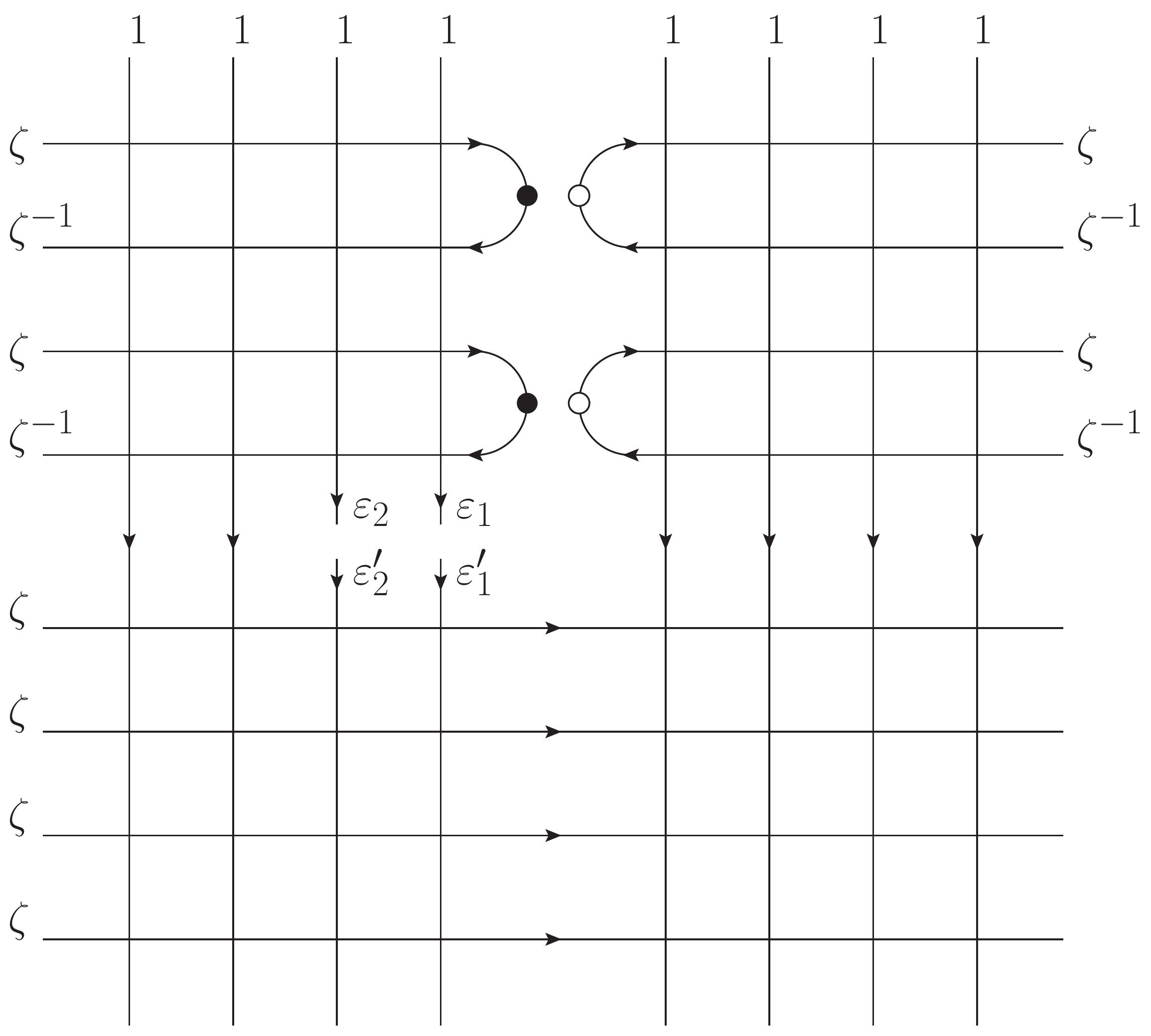}
\caption{The Correlation Function}
\label{fracorr1}
\end{figure}

The other objects of interest for us are correlation functions of the form 
\ben _{(i)}\langle\hb{vac}|E_{\ep'_m}^{\ep_m\, (m)}\cdots E_{\ep'_2}^{\ep_2\,(2)}E_{\ep'_1}^{\ep_1\,(1)}  \vac'_{(i)},\een 
 where $ E_{\ep'}^{\ep\,(j)}$ is the operator that acts at the $j$'th site of the lattice as \bea 
E_{\ep'}^{\ep} (v_{a})=\delta_{a,\ep}
v_{\ep'}.\lb{locop}\eea
We identify this correlation function with the partition function of Figure \ref{fracorr1}  (shown for the case $m=2$), in which we sum  over only those configurations with the indicated spins fixed to the values $(\ep_m,\cdots,\ep_2,\ep_1)$ and $(\ep'_m,\cdots,\ep'_2,\ep'_1)$. We again borrow the
technology of the  vertex operator approach to the bulk and boundary XXZ chains in order to compute these correlation functions. 

\section{The Vertex Operator Approach}
\setcounter{equation}{0}
In this section, we describe the realisation of the objects $\cF^{(i)}$, $T(\z)$, $T'(\z)$ and $E_{\ep'}^{\ep\,(j)}$  within the vertex operator approach. We rely heavily on the previous papers \cite{JM,JKKKM,JKKMW} and refer the reader there for a fuller explanation. 

\subsection{Identification of Spaces and Operators}
It is possible to define an action of the quantum affine algebra $\uq$ on the space $\cH_L^{(i)}$ by making use of the infinite coproduct. The essence of the vertex operator approach is that with respect to this action it is possible to identify $\cH_L^{(i)}$ with the infinite-dimensional irreducible highest-weight $\uq$ module $V(\gL_i)$ - where $\gL_0,\gL_1$ are the fundamental weights. 
Similarly, the vector space $\cH_R^{(i)}\simeq \cH_L^{(i)*}$ can be upgraded to a dual  $\uq$ module (see Chapter 7 of \cite{JM} for the meaning of the $*$ in this case) and is
identified with  $V(\gL_i)^*$. Thus, there is a  $\uq$ identification
\bea \cF^{(i)} \simeq V(\gL_i)\ot  V(\gL_i)^* \simeq \End(V(\gL_i).\lb{spaceident}\eea
The right-hand isomorphism in \mref{spaceident} is given by identifying $a\ot b^*\in  V(\gL)\ot V(\gL_i)^*$ with $ a \times b^* (.)\in \End(V(\gL_i))$.

\begin{figure}[htbp]
\centering
\includegraphics[width=10cm]{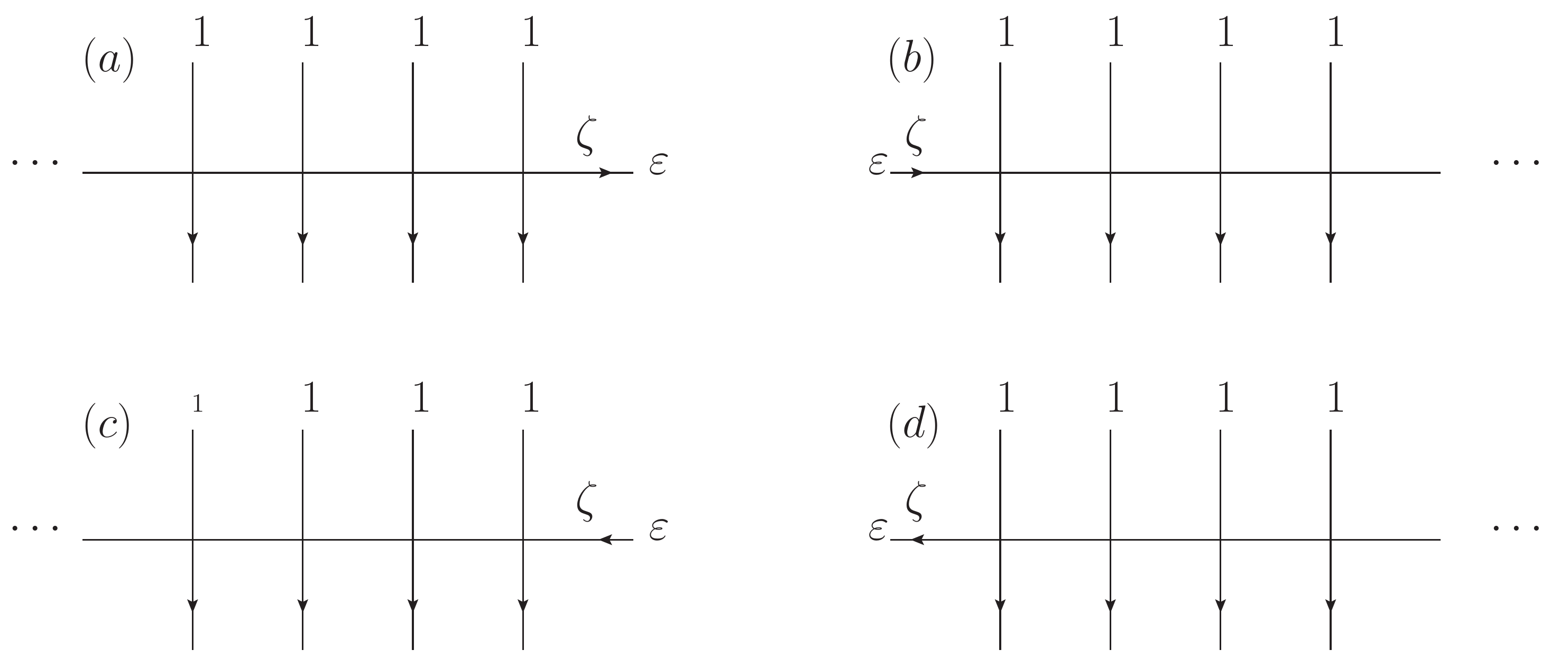}
\caption{The Four Half-Infinite Transfer Matrices}
\label{htms}
\end{figure}

The next step is an identification of the half-infinite lattice transfer matrix which is shown in Figure \ref{htms}(a) and is a map from $\cH_L^{(i)}\ra \cH_L^{(1-i)}$. 
This is identified in terms of a uniquely defined $\uq$ map
\ben \Phi(\z) : V(\gL_i) \ra   V(\gL_{1-i})\ot V_\z,\een
where $V_\z$ is the principal evaluation module associated with the two-dimensional $U_q(sl_2)$ module $V=\C v_+ \oplus \C v_-$ \cite{JM}. Components $\Phi_\pm(\z)$ of this `vertex operator' $\Phi(\z)$ are defined by
\ben \Phi(\z)(a)=\sli_{\ep=+,-} \Phi_\ep(\z)(a) \ot v_{\ep}\een
and are vector-space maps $V(\gL_i) \ra V(\gL_{1-i})$. The associated transpose map $V(\gL_i)^* \ra V(\gL_{1-i})^*$ is denoted by $\Phi_\ep(\z)^t$. 
The left half-infinite transfer matrix of   Figure \ref{htms}(a) is then identified with the operator $\Phi_\ep(\z):V(\gL_i)\ra 
 V(\gL_{1-i})$, whereas the right  half-infinite transfer matrix of   Figure \ref{htms}(b) is identified as $\Phi_{-\ep}(\z)^t: V(\gL_i)^*\ra 
 V(\gL_{1-i})^*$ (both identifications are up to a normalisation factor $g^\half$ defined by Equation \mref{gdef} below). 

A dual vertex operator $\Phi^*(\z):V(\gL_i)\ot V_\z \ra V(\gL_{1-i})$, its components $\Phi^*_\pm(\z)$, and transpose $\Phi^*_\pm(\z)^t$ are defined
 similarly in \cite{JM}. The two remaining half-infinite transfer matrices of Figure \ref{htms}(c) and (d) are then identified as $\Phi^*_\ep(\z):V(\gL_i)\ra V(\gL_{1-i}) $
 and $\Phi^*_{-\ep}(\z)^t:V(\gL_i)^*\ra V(\gL_{1-i})^*$ respectively. In summary, the four half-infinite transfer matrices of Figure \ref{htms} are identified with the four vertex operators
\ben 
&(a)\quad g^{\half}\, \Phi_\ep(\z),\quad &(b)\quad g^{\half}\,\Phi_{-\ep}(\z)^t,\\&(c)\quad g^{\half}\,\Phi^*_\ep(\z),\quad &(d)\quad g^{\half}\,\Phi^*_{-\ep}(\z)^t.\een

The half-infinite transfer matrices of Figure \ref{htms}, and the reflection matrices $K_\bullet(\z)$ and $K_\circ(\z)$  are the building blocks of both $T(\z)$ and $T'(\z)$. Given the identifications of the previous paragraph, it is then straightforward to read off the corresponding realisation of $T(\z)$ and $T'(\z)$ of Figures \ref{bulktrans} and \ref{fractrans} in the vertex operator approach. We have
\bea 
T(\z)&=& g \sli_\ep \Phi_{\ep}(\z)\ot \Phi_{-\ep}(\z)^t\ws:\ws V(\gL_i)\ot V(\gL_i)^* \ra  V(\gL_{1-i})\ot V(\gL_{1-i})^* ,\lb{def:T}\\
T'(\z)&=& T_L(\zeta)\ot T_R(\zeta) \ws:\ws V(\gL_i)\ot V(\gL_i)^* \ra  V(\gL_{i})\ot V(\gL_{i})^*\\[4mm]
\hspace*{-10,mm}\hb{where}\quad 
T_L(\zeta)&=& g \sli_{\ep,\ep'} \Phi_{\ep'}^*(\zeta^{-1}) K_{\!\bullet \ep'}^{\; \ep}(\zeta) \Phi_{\ep}(\z),\lb{tldef}
\\
\hspace*{-10,mm}\hb{and}\quad T_R(\zeta)&=&  g \sli_{\ep,\ep'} \Phi_{-\ep'}^{*}(\zeta^{-1})^t K_{\!\circ \ep}^{\;\ep'}(\zeta) \Phi_{-\ep}(\z)^t.
\eea  
The right transfer matrix 
$T_R(\z)$ can be conveniently rewritten in terms of $T_L(\z)$ by making use of the property \bea
K_\circ(\z)= 
K_\bullet(-q^{-1}\z^{-1})\lb{Kdesfs},\eea and the fact that $\Phi_\ep^*(\z)= \Phi_{-\ep}(-q^{-1} \z)$. In this way, we obtain
\ben T_R(\z)=T_L(-q^{-1}\z^{-1})^t.\een
Furthermore, $T_L(\z)$ defined by \mref{tldef} is simply the same as the boundary transfer matrix of \cite{JKKKM}, there denoted by $T_B(\z)$. It is shown in  \cite{JKKKM} that $T_B(-q^{-1} \z^{-1})=T_B(\z)$. Hence, in summary we have 
\bea T'(\z)&=& T_B(\z)\ot T_B(\zeta)^t,\quad\hb{where}\quad T_B(\zeta)= g \sli_{\ep,\ep'} \Phi_{\ep'}^*(\zeta^{-1}) K_{\!\bullet \ep'}^{\; \ep}(\zeta) \Phi_{\ep}(\z).\lb{tbdef}\eea

\subsection{Eigenstates}
The eigenstates of the bulk transfer matrix $T(\z)$ were first found in the paper \cite{Daval}, and are 
presented in the current notation in \cite{JM}. The form of the eigenstates is neatest if they are written in the space $\End(V(\gL_i))\simeq V(\gL)\ot V(\gL_i)^*$.
The state $\vac_{(i)}$ is particularly simply and is just proportional to $(-q)^D$, where $D$ is a $\uq$ derivation that defines a grading of $V(\gL_i)$ given by $D\ket{\gL_i}=\ket{\gL_i}$ on the highest weight state $\ket{\gL_i}\in V(\gL_i)$ and $D f_{j_1} f_{j_2}\cdots f_{j_m}\ket{\gL_i}=
m \, f_{j_1} f_{j_2}\cdots f_{j_m} \ket{\gL_i}$ on a descendant state produced by the action of $\uq$ generators $f_{0}$ or $f_1$. The inner product of two states in $\End(V(\gL_i))$ is defined in \cite{JM} by $(f,g)=\Tr_{V(\gL_i)}(f\circ g)$, and the 
normalised vacuum state is specified by
\bea 
&&\vac_{(i)}= \frac{1}{\chi^\half} (-q)^D,\quad \hb{with}\quad  
\chi=\Tr_{V(\gL_i)}\big((-q)^{2D}\big)=  \frac{1}{(q^2;q^4)_\infty}, \label{eqn:chi}
\\ &&\hb{for which} \quad   T(\z) \vac_{(i)}= \vac_{(1-i)},\quad \hb{and}\quad _{(i)} 
\dvac| \hb{vac}\rangle_{(i)}=\frac{1}{\chi}\Tr_{V(\gL_i)}\big((-q)^{2D}\big)=1.\nn\eea
We use the standard infinite-product notation \ben (a;b_1,b_2,\cdots,b_N)_\infty=\prod_{n_1=0}^\infty \cdots \prod_{n_N=0}^\infty (1-a \, b_1^{n_1} b_2^{n_2} \cdots  b_N^{n_N}).\een
Also, note that we use the term eigenstate loosely here: $\vac_{(i)}$ is only really an eigenstate of $T^2(\z)$ because of the $i\ra 1-i$ shift above. It is however a genuine eigenstate of the XXZ Hamiltonian $H$ given in Equation \mref{bulkham}.

Vacuum eigenstates $\ket{i}_B\in V(\gL_i)$ of the operator $T_B(\z)$ in the space $V(\gL_i)$ were constructed in the paper \cite{JKKKM}. They were found by solving the eigenstate condition  
\ben T_B(\z)\ket{i}_B = \Lambda^{(i)}(\z) \ket{i}_B,\een where the function $\Lambda^{(i)}(\z)$ is specified in Section 2 of \cite{JKKKM}. This equation is solved by using the free-field realisation of the module  $V(\gL_i)$. The dual vacuum state ${_B}\bra{i}$ was similarly found by solving the condition 
\ben {_B}\bra{i}\,T_B(\z) = \Lambda^{(i)}(\z) \, {_B}\bra{i}.\een
Full details, some of which are restated in Appendix B of the current paper, can be found in \cite{JKKKM}.

In our fractured XXZ model, it is the operator $T'(\z)=T_B(\z)\ot T_B(\z)^t$ that we wish to diagonalise. The form of this expression characterises $T'(\z)$ as an operator on the space $V(\gL_i)\ot V(\gL_i)^*$; the associated action on an element $f\in \End(V(\gL_i))$ is simply 
\ben T'(\z)(f)= T_B(\z) \circ f \circ T_B(\z).\een
Hence, we can immediately write down the normalised vacuum vector $\vac'_{(i)}$ of $T'(\z)$ as 
\ben  \vac'_{(i)}=\frac{1}{ {_B}\langle i \ket{i}_B}\ket{i}_B {_B}\bra{i} \,\in \End(V(\gL_i)),\een
or equivalently as 
\ben  \vac'_{(i)}=\frac{1}{ {_B}\langle i \ket{i}_B}\ket{i}_B \ot  {_B}\bra{i} \,\in V(\gL_i)\ot V(\gL_i)^*.\een
We then have 
\ben T'(\z) \vac'_{(i)}&=& \frac{1}{ {_B}\langle i \ket{i}_B}
 T_B(\z)\, \ket{i}_B {_B}\bra{i}  \, T_B(\z)= (\gL^{(i)}(\z))^2  \vac'_{(i)},\\
\hb{and}\ws {_{(i)}}'\!\langle \hb{vac} \vac'_{(i)}
&=& \frac{1}{\left( {_B}\langle i \ket{i}_B\right)^2}  \Tr_{V(\gL_i}
\Big(\ket{i}_B {_B}\langle i \ket{i}_B {_B}\bra{i}\Big) = 1  .\een

The overlap between the bulk and fractured vacua $ \vac_{(i)}$ and $ \vac'_{(i)}$ is given by the following expression:
\ben {_{(i)}}\!\langle \hb{vac} \vac'_{(i)}&=& \frac{1}{\chi^\half {_B}\langle i \ket{i}_B} \Tr_{V(\gL_i)}\Big((-q)^D \ket{i}_B\, {_B}\bra{i}\Big)=
\frac{1}{\chi^\half {_B}\langle i |i\rangle_B} {_B}\langle i |(-q)^{D}|i\rangle_B.\een
In the next section we use the free-field realisation to compute this overlap as a function of the magnetic-field parameter $r$.

Excited states of the bulk Hamiltonian $H$ are constructed in the vertex operator picture in \cite{JM}. The extra piece of technology required is new `type II' vertex operators defined as suitably normalised $\uq$ intertwiners:
\ben \Psi^*(\xi) : V_\xi\ot V(\gL_i) \ra V(\gL_{1-i}),\quad \Psi(\xi): V(\gL_i) \ra V_\xi \ot V(\gL_{1-i}),\een
 whose components  $\Psi^*_\pm(\xi),\Psi_\pm(\xi) $ are defined by 
\ben \Psi^*(\xi)( v_\pm \ot a)= \Psi^*_\pm (\xi)(a),\quad \Psi(\xi) (a)=\sli_{\ep=\pm} (v_\ep\ot \Psi_\ep(\xi) (a)).\een
The key feature of these new vertex operators is that they quasi-commute with the `type I' vertex operators $\Phi_\pm(\z)$. Namely, 
\bea \Phi_{\ep}(\z) \Psi^*_{\ep'}(\xi) &=& \tau(\z/\xi) \Psi^*_{\ep'}(\xi)\Phi_{\ep}(\z),\lb{com1}\\
\Psi_{\ep'}(\xi)\Phi_{\ep}(\z)&=&  \tau(\z/\xi) \Phi_{\ep}(\z)\Psi_{\ep'}(\xi)\lb{com2},\eea
where $\tau(z)$ is the function 
\ben \tau(z)= \frac{1}{z} \frac{(qz^2;q^4)_\infty (q^3 z^{-2};q^4)_\infty }
{(qz^{-2};q^4)_\infty (q^3z^{2};q^4)_\infty }.\een 
It then follows from \mref{def:T} and \mref{com1} that for $f\in \End(V(\gL_i))$, we have 
\ben T(\z)\big( \psi^*_\ep(\xi) f \big)= \tau(\z/\xi) \, \psi^*_\ep(\xi) \, T(\z) \big(f\big).\een
Thus all states of the form 
\ben \psi_{\ep_1}^*(\xi_1) \cdots \psi_{\ep_{2m}}^*(\xi_{2m}) \vac_{(i)}\in \End(V(\gL_i)) ,\een
are eigenstates of $T(\z)$ with eigenvalue $\tau(\z/\xi_1)\cdots \tau(\z/\xi_{2m})$. These  states are conjectured to span the space $\cF^{(i)}\simeq \End(V(\gL_i))$ in \cite{JM}.

In the boundary case, it follows from  the definition of $T_B(\z)$ given by  Equation \mref{tbdef} and from Equation \mref{com1} that
\bea T_B(\z) \,\psi_{\ep_1}^*(\xi_1) \cdots \psi_{\ep_{2m}}^*(\xi_{2m}) \ket{i}_B
=\gamma(\z;\xi_1,\cdots,\xi_{2m}) \gL^{(i)}(\z) \,\psi_{\ep_1}^*(\xi_1) \cdots \psi_{\ep_{2m}}^*(\xi_{2m}) \ket{i}_B,\lb{lboundcom}\eea
where
$ \gamma(\z;\xi_1,\cdots,\xi_{2m}) =\prod_{n=1}^{2m} \tau(\z \xi_n^{-1}) \, \tau(\z^{-1} \xi_n^{-1})$.
In the context of the fractured model, in addition to \mref{lboundcom}, we also have  
\ben
 {_B}\langle i| \psi_{\ep_{2\ell}}(\xi_{2\ell}) \cdots \psi_{\ep_{1}}(\xi_{1})\, T_B(\z) = \gamma(\z;\xi_1,\cdots,\xi_{2\ell}) \gL^{(i)}(\z)\, 
{_B}\langle i| 
\psi_{\ep_{2\ell}}(\xi_{2\ell}) \cdots \psi_{\ep_{1}}(\xi_{1}).\een
Thus a general excited state is of the form 
\ben \frac{1}{ {_B}\langle i \ket{i}_B}
 \psi_{\ep_1}^*(\xi_1) \cdots \psi_{\ep_{2m}}^*(\xi_{2m})
\ket{i}_B {_B}\bra{i} 
\psi_{\ep'_{2\ell}}(\xi'_{2\ell}) \cdots \psi_{\ep'_{1}}(\xi'_{1})
\een
with $T'(\z)$ eigenvalue given by
\ben \gamma(\z;\xi_1,\cdots,\xi_{2m}) \, \gamma(\z;\xi'_1,\cdots,\xi'_{2\ell})\, (\Lambda^{(i)}(\z))^2.\een

\subsection{Correlation Functions}\lb{sec:cfn}
The realisation of local operators $E_{\ep'}^{\ep\,(j)}$ (defined by Equation \mref{locop}) acting at the $j$'th site of the lattice in terms of type I vertex operators is given in Section 9.1 of \cite{JM}. The product
$E_{\ep'_m}^{\ep_m(m)}\cdots E_{\ep'_2}^{\ep_2(2)}E_{\ep'_1}^{\ep_1(1)}$ is identified with the following operator on $V(\gL_i)\ot V(\gL_i)^*$:
\bea g^m \, \Big(\Phi_{-\ep_1'}(-q^{-1})\Phi_{-\ep_2'}(-q^{-1}) \cdots \Phi_{-\ep_m'}(-q^{-1}) \,
\Phi_{\ep_m}(1) \cdots  \Phi_{\ep_2}(1)  \Phi_{\ep_1}(1)\ot \id\Big) .\lb{resol}\eea
This realisation is the solution of the quantum inverse problem in the vertex operator picture. Let us define (for $N$ even) the following matrix element of vertex operators
\bea
P^{(i)}(\zeta_1,\zeta_2,\cdots,\zeta_N)= \frac{1}{ _{(i)} \langle\hb{vac}\vac'_{(i)}}
{_{(i)}} \langle\hb{vac}|\Phi(\zeta_1)\Phi(\zeta_2) \cdots \Phi(\zeta_N)\otimes \id\vac'_{(i)},\lb{origcf}\eea
in which we are viewing the two vacuum states as a elements of $V(\gL_i)\ot V(\gL_i)^*$ and the product of vertex operators acts on the space $V(\gL_i)$. Using the realisation of the vacua as elements of $\End(V(\gL_i))$ given in the previous subsection yields the equivalent expression
\bea P^{(i)}(\zeta_1,\zeta_2,\cdots,\zeta_N)=\frac{1}{{_B}\langle i | (-q)^{D} |i\rangle_B} {_B}\langle i |(-q)^{D} \Phi(\zeta_1)\Phi(\zeta_2) \cdots \Phi(\zeta_N)|i\rangle_B.\lb{cf2}\eea
or in components: 
\bea P^{(i)}(\zeta_1,\zeta_2,\cdots,\zeta_N)_{\ep_1,\ep_2,\cdots,\ep_m}=\frac{1}{{_B}\langle i | (-q)^{D} |i\rangle_B} {_B}\langle i |(-q)^{D} \Phi_{\ep_1}(\zeta_1)\Phi_{\ep_2}(\zeta_2) \cdots \Phi_{\ep_N}(\zeta_N)|i\rangle_B.\nn\\\lb{cf3}\eea
Using the solution of the quantum inverse problem given by \mref{resol}, we then have the following expression for correlation functions:
\bea &&\hspace*{-5mm}\frac{1}{ _{(i)} \langle\hb{vac}\vac'_{(i)}} _{(i)}\langle\hb{vac}|E_{\ep'_m}^{\ep_m(m)}\cdots E_{\ep'_2}^{\ep_2(2)}E_{\ep'_1}^{\ep_1(1)}  \vac'_{(i)}=
g^m P^{(i)}(\z_1,\z_2,\cdots,\z_{2m})_{-\ep'_1,\cdots,-\ep'_m,\ep_m,\cdots,\ep_1},\nn\\
&&\hspace*{-5mm}\hb{with the choice}\ws \z_1=\z_2=...=\z_m=-q^{-1},\ws \z_{m+1}=\z_{m+2}=...=\z_{2m}=1.\lb{cf4}\eea
A general integral expression for $ P^{(i)}(\zeta_1,\zeta_2,\cdots,\zeta_N)_{\ep_1,\ep_2,\cdots,\ep_N}$ is given in Section \ref{sec:genint} of this paper.

\subsection{The Corner Transfer Matrix Approach}\lb{sec:ctm}
In Section \ref{sec:cfn}, we have expressed correlation functions in terms of vertex operators via \mref{cf4}, where $P^{(i)}(\z_1,\z_2,\cdots,\z_N)$ is defined by \mref{origcf}. Using the definition of $\vac^{(i)}$ and $\vac^{'(i)}$ then gives expression \mref{cf2} for $P^{(i)}(\z_1,\z_2,\cdots,\z_N)$ in terms of the boundary states $\ket{i}_B$ and the operator $(-q)^D$.
While the the argument leading to \mref{cf4} is hopefully clear - at least when supplemented by the further details in \cite{JM,JKKKM} -  the reader might still be lacking a simple, intuitive understanding of why correlation functions can be expressed via \mref{cf3} and \mref{cf4}. Such an understanding is provided by considering corner transfer matrices.

\begin{figure}[htbp]
\centering
\includegraphics[width=5cm]{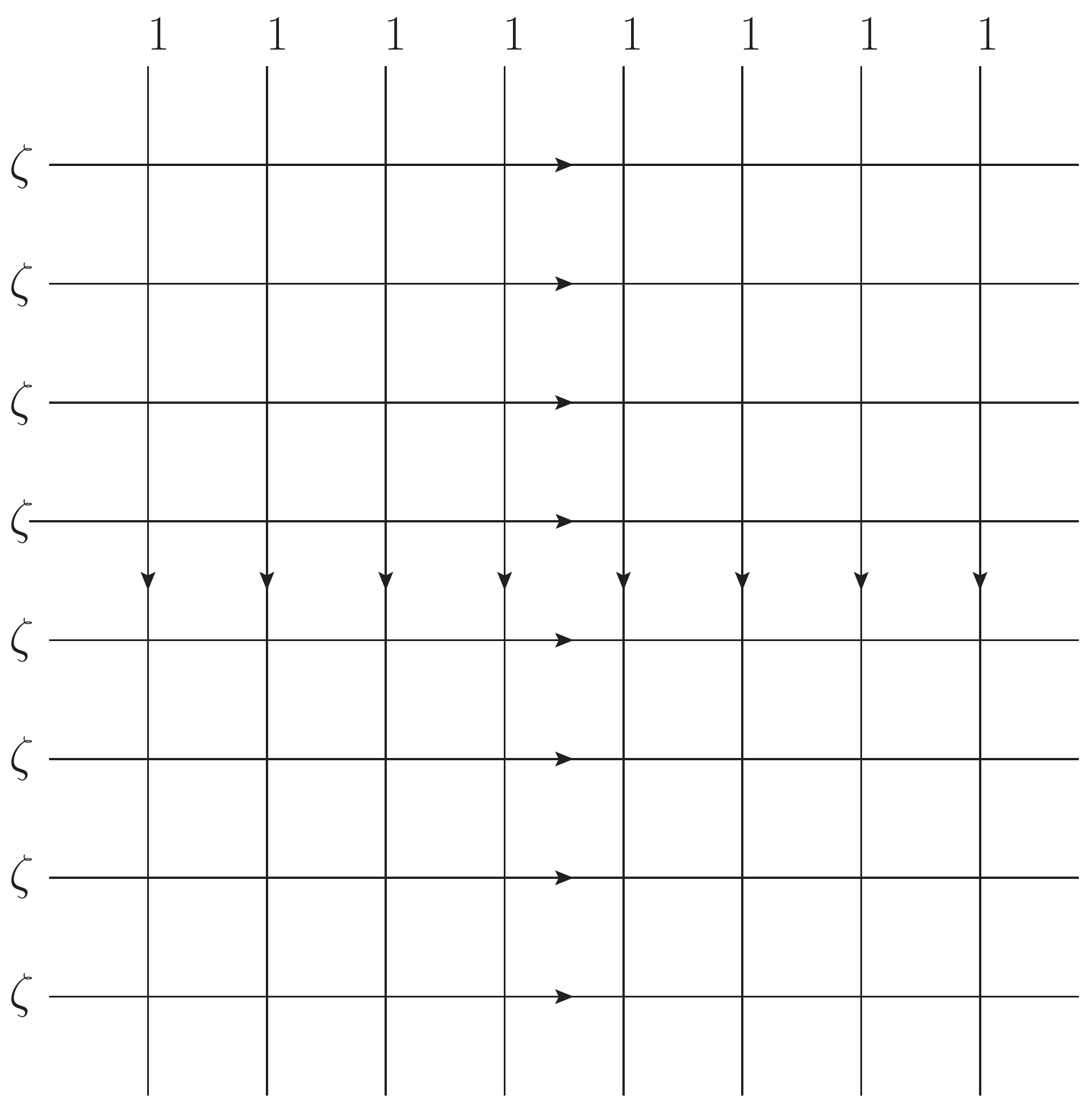}
\caption{The Bulk Partition Function}
\label{bulkpfn}
\end{figure}

The idea of corner transfer matrices was introduced by Baxter, and their role in the vertex operator approach is described for bulk models in \cite{JM} and for boundary models in \cite{JKKMW}. In brief, the 6-vertex model partition function which is associated with the Boltzmann sum over the infinite lattice with antiferromagnetic boundary conditions, and is represented by Figure \ref{bulkpfn}, can be expressed as the trace of the product of four anticlockwise-acting corner transfer matrices (CTMs) $A^{(i)}_{NW}(\z), A^{(i)}_{SW}(\z), A^{(i)}_{SE}(\z), A^{(i)}_{NE}(\z)$ representing the four quadrants of the lattice thus:
\ben Z^{(i)}_{bulk}=\Tr_{\cH^{(i)}_L}
\big( A^{(i)}_{NE}(\z) A^{(i)}_{SE}(\z) A^{(i)}_{SW}(\z)A^{(i)}_{NW}(\z)\big).\een

\nin The CTMs act on the half-spaces $\cH^{(i)}_L$ or $\cH^{(i)}_R$ as follows: 
\ben  A^{(i)}_{SW}(\z): \cH^{(i)}_L \ra  \cH^{(i)}_L,\quad  A^{(i)}_{SE}(\z): \cH^{(i)}_L \ra  \cH^{(i)}_R,\quad A^{(i)}_{NE}(\z): \cH^{(i)}_R \ra \cH^{(i)}_R,\quad A^{(i)}_{NE}(\z):\cH^{(i)}_R\ra \cH^{(i)}_L.\een 
The remarkable property of corner transfer matrices discovered by Baxter is that in the infinite lattice limit we have $A(\z):=A^{(i)}_{SW}(\z) \sim \z^{-D}$, where $D:\cH_{L}^{(i)}\ra \cH_{L}^{(i)}$ is an operator with non-negative integer eigenvalues and 
where $\sim$ indicates equality up to a normalisation factor in this limit. We use the same notation $D$ as for the $\uq$ derivation because we shall identify the two objects. The crossing symmetry of the 6-vertex R-matrix given by \mref{xing} then allows us to relate the other three CTMs to $A(\z)$ thus: $A^{(i)}_{SE}(\z)= S A(-q^{-1}\z^{-1})$, $A^{(i)}_{NE}(\z)= S A(\z) S$, $A^{(i)}_{NW}(\z)= A(-q^{-1}\z^{-1}) S$, where $S= \cdots \sigma^x\ot \sigma^x$. Hence, we have
\ben Z_{bulk}^{(i)}\sim \Tr_{ \cH^{(i)}_L}   \big((-q)^{2D}\big).\een

\begin{figure}[htbp]
\centering
\includegraphics[width=3cm]{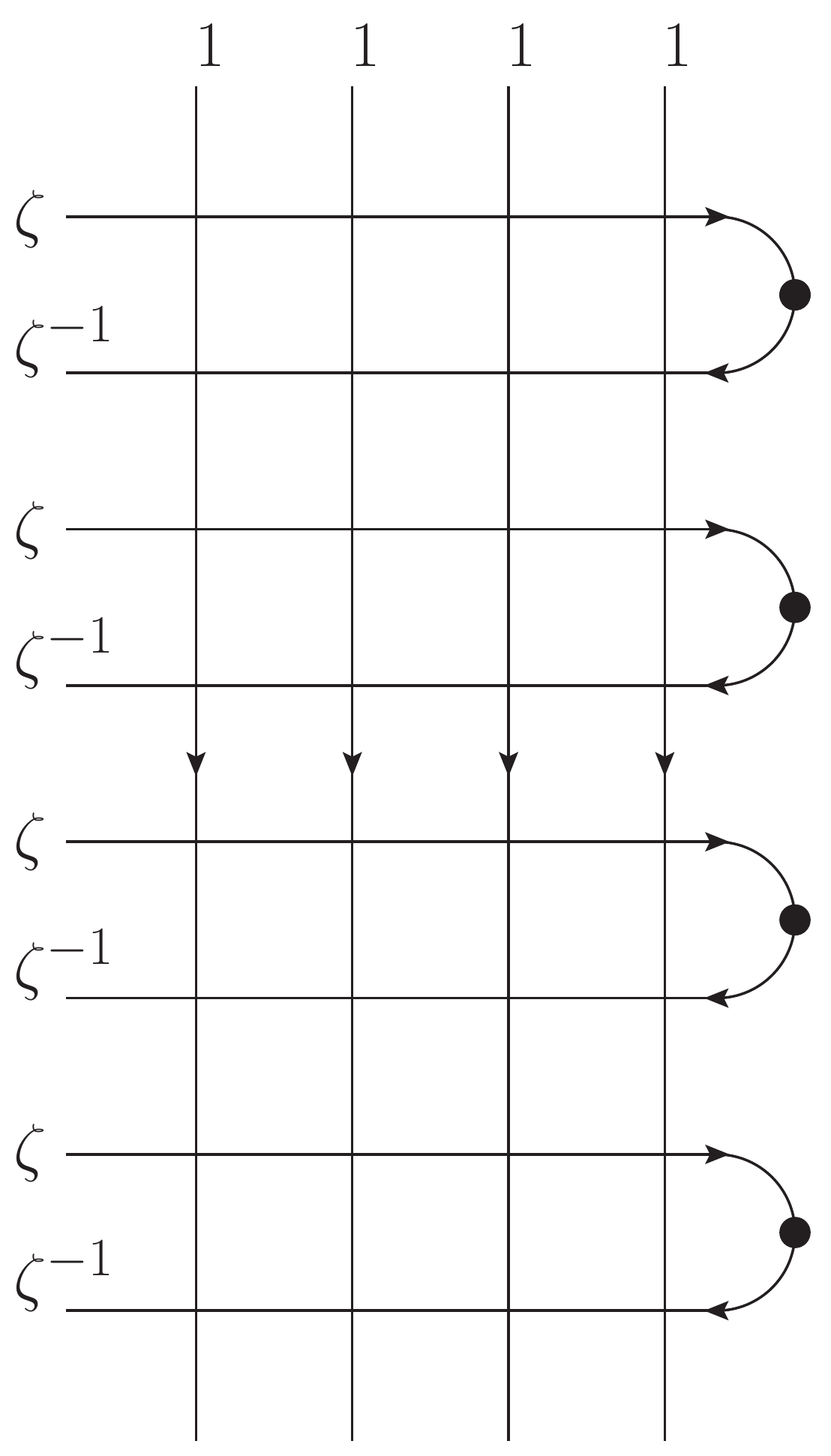}
\caption{The Boundary Partition Function}
\label{boundpfn}
\end{figure}

The boundary partition function $Z^{(i)}_{boundary}$ considered in \cite{JKKKM,JKKMW} is represented by Figure \ref{boundpfn}. 
The modified CTMs associated with this model were considered in detail in \cite{JKKMW}. The NW and SW CTMs of Figure \ref{boundpfn} were there labelled $A^{(i)}_{NW}(\z,1)$ and $A^{(i)}_{SW}(\z,1)$, and the upper-boundary and dual lower-boundary states were denoted by $\ket{B;\z}^{(i)}$ and $^{(i)}\bra{B;\z}$ (the implicit $\z$ dependence was suppressed in \cite{JKKMW}). In this boundary case we therefore have
\ben Z^{(i)}_{boundary} = {^{(i)}}\langle B;\z| A^{(i)}_{SW}(\z,1) A^{(i)}_{NW}(\z,1) \ket{B;\z}^{(i)}.\een
Identifying the boundary states as $\ket{i}_B \sim A^{(i)}_{NW}(\z,1)\ket{B;\z}^{(i)}$, ${_B}\langle i|\sim {^{(i)}}\langle B;\z|A^{(i)}_{SW}(\z,1)$  then leads to
\ben Z^{(i)}_{boundary} \sim {_B}\langle i\ket{i}_B.\een

The partition function of the fractured model shown in Figure \ref{fracpfn} is built by combining the CTMs and states of both the bulk and boundary models. It is apparent from Figure \ref{fracpfn} that we have 
\ben Z^{(i)}_{fracture}= {^{(i)}}\!\!\!{_\circ}\langle B;\z|  A^{(i)}_{NE}(\z,1)   A^{(i)}_{SE}(\z)  A^{(i)}_{SW}(\z) A^{(i)}_{NW}(\z,1) \ket{B;\z}^{(i)}_\bullet,\een
where ${^{(i)}}\!\!\!{_\circ}\langle B;\z|$ denotes the dual boundary state on the right of the fracture in Figure \ref{fracpfn}, $A^{(i)}_{NE}(\z,1)$ denotes the NE corner transfer matrix in the same figure, and $\ket{B;\z}^{(i)}_\bullet=\ket{B;\z}^{(i)}$. Using the crossing symmetry \mref{xing} of the R-matrix and the relationship between $K_\circ$ and $K_\bullet$ given by \mref{Kdesfs}, it then simple to show that we can identify
\ben {^{(i)}}\!\!\!{_\circ}\langle B;\z|        A^{(i)}_{NE}(\z,1) = {^{(i)}}\langle B;-q^{-1}\z^{-1}| A^{(i)}_{SW}(-q^{-1}\z^{-1},1) S\sim {_B}\langle i| S.\een
Hence we have
\ben Z^{(i)}_{fracture}= {^{(i)}}\langle B;\z| A^{(i)}_{SW}(\z,1)   A^{(i)}(-q^{-1}\z^{-1})  A^{(i)}(\z) A^{(i)}_{NW}(\z,1) \ket{B;\z}^{(i)}
\sim  {_B}\langle i |(-q)^D    \ket{i}_B .\een

Correlation functions associated with horizontal sites $m,\cdots,2,1$ labelled as in Figure \ref{fracorr1}, can then be constructed in by inserting a
product of $2m$ half-infinite transfer matrices of type (a) of Figure \ref{htms} beneath the NW CTM in any of the
bulk, boundary or fractured scenarios above (the particular product of half-infinite transfer matrices required is always that of our expression \mref{resol} - a fact which
again comes from the solution of the quantum inverse problem described in \cite{JM}). Identifying these 
half-infinite transfer matrices with the vertex operators $\Phi_\ep(\z)$, and $H_L^{(i)}$ with $V(\gL_i)$,  leads immediately to the general form for correlation functions (with $N$ even) 
\bea  F^{(i)}(\z_1,\z_2,\cdots,\z_N) &=& \frac{1}{\Tr_{V(\gL_i}\big((-q)^{2D}\big)} 
 \Tr_{V(\gL_i}\big((-q)^{2D}\Phi(\z_1)\Phi(\z_2)\cdots \Phi(\z_N)\big), \lb{bulkcof}\\
 G^{(i)}(\z_1,\z_2,\cdots,\z_N)&=&\frac{1}{ {_B}\langle i\ket{i}_B } 
 {_B}\langle i|\Phi_{\ep_1}(\z_1)\Phi_{\ep_2}(\z_2)\cdots \Phi_{\ep_N}(\z_N)\ket{i}_B,\lb{boundcof} \\
 P^{(i)}(\z_1,\z_2,\cdots,\z_N)&=&\frac{1}{ {_B}\langle i| (-q)^D \ket{i}_B } 
 {_B}\langle i| (-q)^D \Phi_{\ep_1}(\z_1)\Phi_{\ep_2}(\z_2)\cdots \Phi_{\ep_N}(\z_N)\ket{i}_B,\lb{fraccof}
\eea
found in the bulk, boundary and fractured models respectively.

\subsection{The Boundary qKZ Equation}
Boundary qKZ equations first appeared in the work of Cherednik \cite{Cher} and were realised in a boundary lattice model in \cite{JKKMW}. 
The two functions \mref{boundcof} and \mref{fraccof} obey boundary qKZ equations of different levels (we use the level terminology of \cite{MR2186202,MR2367185}). This follows from the the following properties:
\bea
K(\z) \Phi(\z)\ket{i}_B&=& \gL^{(i)}(\z;r) \phi(\z^{-1})\ket{i}_B,\lb{pr1}\\
\hat{K}(-q^{-1}\z)\; {_B}\langle i| \Phi(\z^{-1})&=& \gL^{(i)}(-q^{-1}\z;r) {_B}\langle i| \Phi(q^{-2}\z),\lb{pr2}\\
\hat{K}(q^{-2}\z) \; {_B}\langle i| (-q)^D \Phi(\z^{-1})&=& \gL^{(i)}(q^{-2}\z;r) {_B}\langle i| (-q)^D\Phi(q^{-4}\z)\lb{pr3},\\
PR(\z_1/\z_2) \Phi(z_1) \Phi(\z_2) &=& \Phi(\z_2) \Phi(\z_1)\lb{pr4},
\eea
where the matrix $\hat{K}(\z)$ is defined by $\hat{K}^{\ep'}_\ep(\z)=K^{-\ep}_{-\ep'}(\z)$.
The first two equalities \mref{pr1} and \mref{pr2} follow from the defining properties of $|i\rangle_B$ and $_{B}\!\langle i|$: 
\ben T_B(\z)\, |i\rangle_B=\Lambda^{(i)}(\z)\,|i\rangle_B,\quad\hb{and}\quad  _{B}\!\langle i|\, T_B(\z)= _{B}\!\langle i|\,\Lambda^{(i)}(\z;r)\een 
together with the following properties of vertex operators \cite{JM}:
\ben g\sli_{\ep} \Phi^*_\ep(\z) \Phi_\ep(\z)=\id,\quad g\, \Phi_{\ep}(\z) \Phi^*_{\ep'}= \delta_{\ep,\ep'}\id,\quad 
\Phi^*_\ep(\z)=\Phi_{-\ep}(-q^{-1}\z).\een
Equation \mref{pr3} then follows from \mref{pr2} supplemented by the vertex operator property $(-q)^D \Phi(\z)= \Phi(-q\z) (-q)^D$. The final relation \mref{pr4} is the standard intertwining relation for vertex operators where $P$ is the permutation operator $P(v_{\ep_1}\ot v_{\ep_2})=v_{\ep_2}\ot v_{\ep_1}$.

The boundary qKZ relations are simply the relations for the functions $G^{(i)}(\z_1,\z_2,\cdots,\z_N)$ or $P^{(i)}(\z_1,\z_2,\cdots,\z_N)$ above that follow from \mref{pr1}-\mref{pr4}. In the case $i=0$ we have $\gL^{(0)}(\z;r)=1$ \cite{JKKKM} and hence obtain
\bea
&&G^{(0)}(\z_1,\cdots, \z_{j-1},q^{-2} \z_j, \z_{j+1},\cdots,\z_N)=\nn\\
 && R_{j,j-1}(q^{-2}\z_j/\z_{j-1}) \cdots R_{j,1}(q^{-2}\z_j/\z_1) \hat{K}_{j}(-q^{-1}\z_j)\nn\\
\times && R_{1,j}(\z_1 \z_j) \cdots R_{j-1,j}(\z_{j-1}\z_j) R_{j+1,j}(\z_{j+1} \z_j) \cdots R_{n,j}(\z_n\z_j)\nn\\
\times &&K_{j} (\z_j)R_{j,N}(\z_j/\z_N)\cdots R_{j,j+1}(\z_j/\z_{j+1})  G^{(0)}(\z_1,\z_2,\cdots,\z_N),\lb{qKZG}\eea
and 
\bea
&& P^{(0)}(\z_1,\cdots, \z_{j-1},q^{-4} \z_j, \z_{j+1},\cdots,\z_N)=\nn\\
 && R_{j,j-1}(q^{-4}\z_j/\z_{j-1}) \cdots R_{j,1}(q^{-4}\z_j/\z_1)  \hat{K}_{j}(q^{-2}\z_j)\nn\\
\times &&R_{1,j}(\z_1 \z_j) \cdots R_{j-1,j}(\z_{j-1}\z_j) R_{j+1,j}(\z_{j+1} \z_j) \cdots R_{n,j}(\z_n\z_j)\nn\\
\times &&K_{j} (\z_j)R_{j,N}(\z_j/\z_N)\cdots R_{j,j+1}(\z_j/\z_{j+1})  P^{(i)}(\z_1,\z_2,\cdots,\z_N).\lb{qKZP}\eea
Such generalised boundary qKZ equations were considered in \cite{MR2186202}. In the language of \cite{MR2186202} a boundary qKZ equation for a function $\Psi(\z_1,\z_2,\cdots,\z_N)$ is characterised as being of type $(r,s)$ if we have  $\Psi(\z_1,\z_2,\cdots,\z_N)$ related to both $\Psi(\z_1,\z_2,\cdots,r^\half s^\half/\z_N)$ and $\Psi(r^{\half}/\z_1,\z_2,\cdots,\z_N)$, as a consequence of which $\Psi(\z_1,\z_2,\cdots,\z_N)$ is then related to 
$\Psi(\z_1,\z_2,\cdots,s^{-\half}\z_j,\z_N)$.\footnote{The slight difference in notation compared with \cite{MR2186202} arises because $z$ of \cite{MR2186202} is equal to our $\z^2$.} Hence, we have $(r,s)=(q^{-4},q^4)$ for $G^{(0)}$, and  $(r,s)= (q^{-8},q^8)$ for $P^{(0)}$. The `level' $\ell$ of the boundary qKZ is then identified in \cite{MR2186202} as $s=q^{2(k+\ell)}$ (with $k=2$ for the $sl_2$ case) in analogy with the bulk qKZ equation of Frenkel and Reshetikhin \cite{FR}. Hence, in this language, Equation \mref{qKZG} (which appeared in \cite{JKKMW}) is level 0, and \mref{qKZP} is level 2.

\section{An Integral Formula for Correlation Functions}\label{sec:genint}\setcounter{equation}{0}
In this section, we present a general integral formula for $P^{(i)}(\z_1,\z_2,\cdots,\z_N)$ which we then specialise to give the fracture magnetisation
\ben \frac{ _{(i)}\!\langle \hb{vac}| \sigma^z_1 \vac'_{(i)} }{_{(i)}\!\langle \hb{vac}\vac'_{(i)}}  = g \Big( P^{(i)}(-q^{-1},1)_{-+}- P^{(i)}(-q^{-1},1)_{+-}\Big).\een We make use of the free-field realisation described in detail in the paper \cite{JKKKM} and the book \cite{JM}. Most calculational details of the current work are relegated to our Appendix B to which to refer as necessary.   

\subsection{The Matrix Element $ {_{(i)}}\!\langle \hb{vac} \vac'_{(i)}$   }
The boundary states are given in terms of highest-weight states $\ket{\gL_i}$ by
 $|i\rangle_B=e^{F_i} \ket{\gL_i}$ and  ${_B}\langle i|=\langle \gL_i| e^{G_i}$, where $F_i$ and $G_i$ have the following quadratic
form in terms of the q-bosonic oscillators $a_{n}$ ($n\in \Z/\{0\}$):
\ben F_i =\frac{1}{2}\sli_{n=1}^\infty \frac{n \alpha_n}{[2n][n]}a_{-n}^2+ \sli_{n=1}^\infty \beta_n^{(i)} a_{-n},\quad
G_i =\frac{1}{2}\sli_{n=1}^\infty \frac{n \gamma_n}{[2n][n]}a_{n}^2+ \sli_{n=1}^\infty \delta_n^{(i)} a_{n} .\een
The coefficients $\alpha_n,\beta_n^{(i)},\gamma_n, \delta_n^{(i)}$ are functions of the magnetic-field parameter $r$ entering the matrices $K(\z,r)$ of Equation \mref{rm}. They were defined in \cite{JKKKM} and are reproduced in our Appendix B. 

We show in Appendix B that we have
\ben _B \bra{0}(-q)^D |0\rangle_B\hspace*{-2mm}&=&\hspace*{-2mm}\frac{(r^2 q^{10};q^8,q^8)^2_\infty}{ (r^2 q^{12};q^8,q^8)^2_\infty} 
\frac{(r^2 q^{6};q^4,q^8)_\infty}{(r^2 q^{4};q^4,q^8)_\infty} 
\frac{( q^{14} ;q^8,q^8)_\infty}{( q^{10};q^8,q^8)_\infty},\ws  _B \bra{0}0\rangle_B= \frac{(q^4 r^2;q^8)_\infty}{(q^2 r^2;q^8)_\infty (q^6;q^8)_\infty},\een
and that $_B \bra{1}(-q)^D |1\rangle_B$ and $ _B \bra{1}1\rangle_B$ are given by the same expressions with the substitution $r\ra r^{-1}$. 
Hence we obtain 
\bea \hspace*{-7mm}{_{(0)}}\langle \hb{vac} \vac'_{(0)}&=&
\frac{1}{\chi^\half {_B}\langle i |i\rangle_B} {_B}\langle i |(-q)^{D}|i\rangle_B\nn
\\
&=& (q^2;q^4)_\infty^\half
\frac{(r^2 q^{10};q^8,q^8)^2_\infty}{  (r^2 q^{4};q^8,q^8)_\infty (r^2 q^{12};q^8,q^8)_\infty} 
\frac{(r^2 q^{2};q^4,q^8)_\infty}{(r^2 q^{4};q^4,q^8)_\infty}
\frac{( q^{6} ;q^8,q^8)_\infty}{( q^{10};q^8,q^8)_\infty}.\lb{ovlap}\eea
The expression for ${_{(1)}}\langle \hb{vac} \vac'_{(1)}$ is again given by the substitution $r \ra r^{-1}$.  

\begin{figure}[htbp]
\centering
\includegraphics[width=7cm]{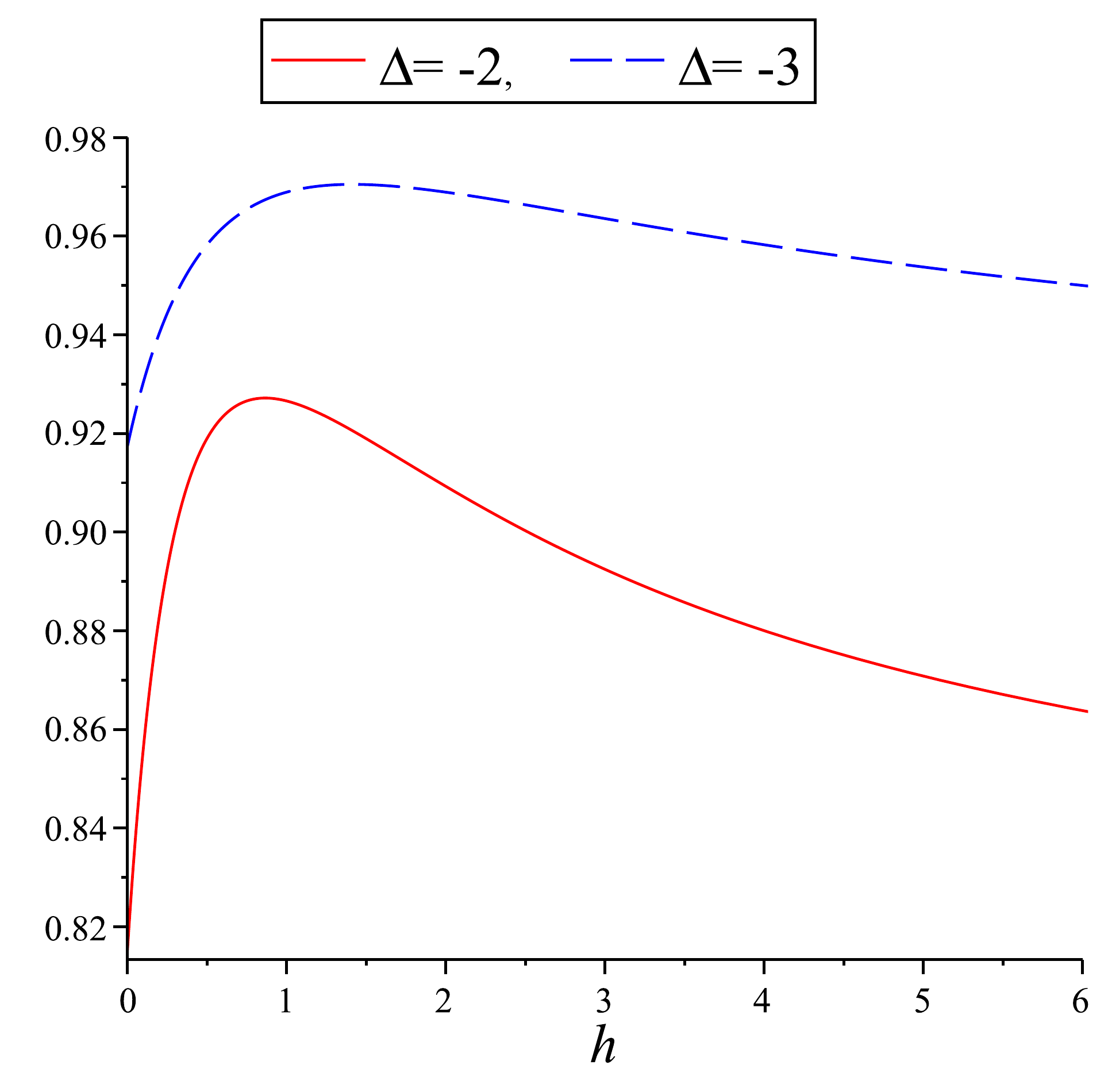}
\caption{The Matrix Element $|{_{(0)}}\langle \hb{vac} \vac'_{(0)}|^2$ as a Function of Magnetic Field $h$}
\label{overlap}
\end{figure}

The matrix element ${_{(0)}}\langle \hb{vac} \vac'_{(0)}$ represents the overlap between the bulk and fractured vacua, and it is of interest to see how this overlap varies as a 
function of the magnetic field $h$ given by \mref{magfield}, for $0\leq h\leq \infty$ corresponding to $-1\leq r\leq 1$. The function ${_{(0)}}\langle \hb{vac} \vac'_{(0)}$ is clearly a symmetric function of $r$ with a maximum at $r=0$. The value of the fracture magnetic field at this point is $h=h_{inv}:=\frac{(q^2-1)}{4q}$ which corresponds to the 
$U_q(sl_2)$ invariant point \cite{MR926391,MR1043392} of the finite boundary Hamiltonian \mref{hfinbound}.
It is interesting that the vacua of $H$ and $H'$ are most similar at this very special value $h_{inv}$ of the magnetic field. 
The quantity $|{_{(0)}}\!\langle \hb{vac} \vac'_{(0)}|^2$ is called the fidelity in the language of quantum information theory, and a graph of the fidelity against $h$ is shown in Figure \ref{overlap} for two different values of the Hamiltonian anisotropy $\Delta=(q+q^{-1})/2$.
The fidelity increases with increasing $|\Delta|$ as can be seen.\\[3mm]

\subsection{The General Integral Expression}
In Appendix B, we consider the free-field realisation of the general, N even, correlation function 
\ben P^{(i)}(\zeta_1,\zeta_2,\cdots,\zeta_N)_{\ep_1,\ep_2,\cdots,\ep_N}=\frac{1}{{_B}\langle i | (-q)^{D} |i\rangle_B} {_B}\langle i |(-q)^{D} \Phi_{\ep_1}(\zeta_1)\Phi_{\ep_2}(\zeta_2) \cdots \Phi_{\ep_N}(\zeta_N)|i\rangle_B,\een
for which we derive an integral expression. 
Letting $A=\{j|1\leq j\leq N, \ep_j=+1\}$, and defining $z_j=\z_j^2$, we find
\bea P^{(i)}(\z_1,\z_2,\cdots,\z_N)_{\ep_1,\ep_2,\cdots,\ep_N}\hspace*{-3mm}&=& \hspace*{-3mm}(-q^3)^{N^2/4+iN/2-\sli_{a\in A } a}(1-q^2)^{N/2} \pl_{j=1}^N \z_j^{\frac{1+\ep_j}{2}-j+N+i} \pl_{j<k} 
\frac{(q^2z_k/z_j;q^4)_\infty}{(q^4 z_k/z_j;q^4)_\infty}\nn\\
&&\hspace*{-25mm} \times\pl_{a\in A} \oint_{C_a} \frac{dw_a}{2\pi\sqrt{-1}} w_a^{1-i} \frac{\pl_{a<b} (w_a-w_b)(w_a-q^2 w_b) }
{\pl_{j\leq a} (z_j-q^{-2} w_a) \pl_{a\leq j} (w_a-q^4 z_j)  } I^{'(i)}(\{z_j\},\{w_a\}),\lb{gencfn}\eea
where $I^{'(i)}(\{z_j\},\{w_a\})$ is given by Equation \mref{cfn8}. The closed contour $C_a$ is an anticlockwise one in the complex plane chosen such that the poles at $q^4 z_j$ lie inside and the poles at $q^2 z_j$ lie outside the contour. Further restrictions on the contour associated with the poles arising from $I^{'(i)}(\{z_j\},\{w_a\})$ are given in Appendix B. Correlation functions are then given by the specialisation corresponding to Equation \mref{cf4} where the constant $g$ is defined in \cite{JM} as 
\bea g=\frac{(q^2;q^4)_\infty}{(q^4;q^4)_\infty}.\label{gdef}\eea
\subsection{The Fracture Magnetisation}
We consider the magnetisation 
\ben M^{(i)}(r):&=&\frac{_{(i)}\langle\hb{vac}|\sigma^z_1 \vac'_{(i)}}{_{(i)}\langle\hb{vac} \vac'_{(i)}}= g \left(P^{(i)}(-q^{-1},1)_{-+}-P^{(i)}(-q^{-1},1)_{+-}\right).\een 
Specialising formula \mref{gencfn} to $N=2$, with $z=\z^2$, we find
\bea g P^{(i)}(-q^{-1}\z,\z)_{-+}= (q^2 z)^i
z (1-q^2)^2\oint_{C^{(i)}_{-+}} \frac{dw}{2\pi \sqrt{-1}} \frac{w^{1-i}}{(w-z)(w-q^2z)(w-q^4z) } I^{'(i)},\lb{pmp}\eea
\ben \hb{where}\quad \quad I^{'(i)}&=&  F^{'(i)}\, (q^8z^2 ;q^8)_\infty (q^4/z^2;q^8)_\infty (q^8;q^8)_\infty(q^{10};q^8)^2_\infty \Theta_{q^8}(q^2w^2) \\
&&\times \frac{1}{ (q^6zw;q^8)_\infty (q^4/(zw);q^8)_\infty (q^{12}z/w;q^8)_\infty (q^6w/z;q^8)_\infty}\\
&&\times \frac{1}{ (q^4zw;q^8)_\infty (q^6/(zw);q^8)_\infty (q^{10}z/w;q^8)_\infty (q^8w/z;q^8)_\infty},
\een
\ben\hb{and} \quad F^{'(0)}&=&\frac{(q^2rz;q^8)_\infty(q^4r/z;q^8)_\infty }
{(q^8rz;q^8)_\infty (q^2r/z;q^8)_\infty  } 
\frac{ (q^6rw;q^8)_\infty  (q^4r/w;q^8)_\infty }{(rw;q^8)_\infty (q^6r/w;q^8)_\infty},\\
F^{'(1)}&=&\frac{(1/(rz);q^8)_\infty(q^6z/r;q^8)_\infty }
{(q^6/(rz);q^8)_\infty (q^4z/r;q^8)_\infty  } 
\frac{ (q^2w/r;q^8)_\infty  (q^8/(rw);q^8)_\infty }{(q^4w/r;q^8)_\infty (q^2/(rw);q^8)_\infty}.\\
\een
The contour $C_{-+}^{(i)}$ is fixed following the prescription discussed in the previous subsection and in further detail in Appendix B. Let us use the notation $ y_1 <w< y_2$ to indicate a $w$ contour which passes outside the $y_1$ pole and inside the $y_2$ pole. Then, to be explicit, the contour is given by the following requirements for all $n\geq 0$:
\bea
C^{(0)}_{-+}:&& q^{6+8n}r, q^{4+8n}/z, q^{6+8n}/z,q^{10+8n}z, q^{12+8n}z, q^4z <w,\nn\\
&&w <q^{-8n}/r, q^{-6-8n}/z, q^{-4-8n}/z,q^{8-8n}z, q^{-6-8n}z, z, zq^2,\nn\\
 C^{(1)}_{-+}:&& q^{2+8n}/r, q^{4+8n}/z, q^{6+8n}/z,q^{10+8n}z, q^{12+8n}z, q^4z <w,\nn\\
&&w <q^{-4-8n}r, q^{-6-8n}/z, q^{-4-8n}/z,q^{8-8n}z, q^{-6-8n}z, z, zq^2.\lb{maincontour}\eea

Similarly, we find
\ben g P^{(i)}(-q^{-1}\z,\z)_{+-}= -(q^2 z)^i
z (1-q^{2})^2 \oint_{C^{(i)}_{+-}} \frac{dw}{2\pi \sqrt{-1}} \frac{w^{1-i}}{(w-z)(w-q^2z)(w-q^4z) } I^{'(i)}\een
where $C^{(i)}_{+-}$ is the same as  $C^{(i)}_{-+}$ except that the pole at $q^2 z$ now lies inside the contour.

A useful check of our formalism is provided by computing the matrix element of the identity operator
$\id^{(1)}=E^{+(1)}_+ + E^{-(1)}_-$. We have 
\ben &&\frac{_{(i)}\langle\hb{vac}|\id^{(1)} \vac'_{(i)}}{_{(i)}\langle\hb{vac} \vac'_{(i)}}= g \left(P^{(i)}(-q^{-1},1)_{-+}+P^{(i)}(-q^{-1},1)_{+-}\right).\een 
We see from the above discussion that 
\ben  g \left(P^{(i)}(-q^{-1},1)_{-+}+P^{(i)}(-q^{-1},1)_{+-}\right) = -
(q^2 z)^i
z (1-q^{2})^2 \hb{Res}_{w={zq^2}}  \left[ \frac{w^{1-i}}{(w-z)(w-q^2z)(w-q^4z) } I^{'(i)}\right].\een
We find that this expression has the required value of 1 for both $i=0$ and $i=1$. 

It remains to compute $P^{(i)}(-q^{-1},1)_{+-}$ . The obvious way to carry out this integral is by a brutal summation of the infinite sets of poles lying inside $C^{(i)}_{+-}$. Another way is to recall the origin of the $w$ integrals in the expression for $P^{(i)}(\zeta_1,\zeta_2,\cdots,\zeta_N)_{\ep_1,\ep_2,\cdots,\ep_N}$: they arise because the vertex operator $\Phi_+(\z)$ is expressed as the (-1)th coefficient of a Laurent expansion in $w$ (see Chapter 8 of \cite{JM}).
 At any given order in $q$, there are only a finite number of contributions to this coefficient that come from
expanding  the numerator and denominator of the integrand of $P^{(i)}(\zeta_1,\zeta_2,\cdots,\zeta_N)_{\ep_1,\ep_2,\cdots,\ep_N}$ (taking care to respect the analyticity requirements expressed in the specified 
contour). Hence, the q-expansion of the correlation function is relatively easy to compute from the integral expression. In this way, we have computed $P^{(i)}(-q^{-1},1)_{+-}$ up to 
order $q^{96}$. The result is consistent with the following conjectural form:
\ben g P^{(0)}(-q^{-1},1;r)_{+-}=(1-r) \sli_{n=0}^{\infty} \frac{(-q^2)^n}{(1-rq^{4n})} ,\een
and also with the required symmetry \bea P^{(i)}(-q^{-1},1;r)_{+-}= P^{(1-i)}(-q^{-1},1;r^{-1})_{-+}\lb{sym1}\eea
(we are now showing the implicit $r$ dependence of the correlation function). 
The sum can be represented as a basic hypergeometric function (see, for example, Chapter 17 of \cite{NIST}) in the following way:
\ben g P^{(0)}(-q^{-1},1;r)_{+-}
=\fullhhg{q^4}{r}{rq^4}{q^4}{-q^2}.\een
Hence we obtain the magnetisation
\bea -M^{(0)}(r)&=& g\Big( -P^{(i)}(-q^{-1}\z,\z;r)_{-+}+P^{(i)}(-q^{-1}\z,\z;r)_{+-}\Big)=-1+2g P^{(0)}(-q^{-1},1;r)_{+-}\nn\\
&=& 1+ 2(1-r) \sli_{n=1}^{\infty} \frac{(-q^2)^n}{(1-rq^{4n})}\lb{magform}\eea
which can also be re-expressed at 
\ben -M^{(0)}(r)&=&2 \, \fullhhg{q^4}{r}{rq^4}{q^4}{-q^2}-1 .\een
It also follows from the symmetry \mref{sym1} that we have \ben -M^{(0)}(r)=M^{(1)}(r^{-1}).\een

The form \mref{magform} is similar, but different, to the magnetisation at site 1 of the pure boundary model of Figure \ref{boundpfn}
 considered in \cite{JKKKM}:
\bea -M^{(0)}_{bound}(r)&=&1+ 2(1-r)^2 \sli_{n=1}^{\infty} \frac{(-q^2)^n}{(1-rq^{2n})^2}\lb{boundmag}.\eea

The three special cases $r=-1,0,1$ corresponding to $h=0,h_{inv},\infty$\,  are of separate interest: 

\vspace*{2mm}

\nin $\bullet$ When $r=-1$ ($h=0$), we can sum the expression \mref{magform} to obtain
\ben 
 -M^{(0)}(r=-1)&=& \frac{(q^2;q^2)_\infty^2}{(-q^2;q^2)_\infty^2}.\een
This coincides with the expression for the spontaneous magnetisation of the bulk model \cite{Bax82} (the minus sign is just due to the choice of convention for the $i=0$ ground state of Figure \ref{fracgs}). It seems physically reasonable that the magnetisation at position 1 in Figure \ref{fracorr1} is simply the same as that of the zero-field  bulk model when the 
fracture magnetic field is zero.

\vspace*{2mm}

\nin $\bullet$ When $r=0$ ($h=h_{inv}$), we have 
\ben -M^{(0)}(r=0)= \frac{1-q^2}{1+q^2}\een
which coincides with the magnetisation \mref{boundmag} of the boundary model at this value of the magnetic field. 
At this point  the overlap ${_{(0)}}\!\langle \hb{vac} \vac'_{(0)}$ is a maximum and thus the bulk system is closest to the tensor product of right and left boundary systems. It is therefore reasonable that the fracture magnetisation at site 1 should coincide with magnetisation at the corresponding site of the left boundary system shown in Figure \ref{boundpfn}.

\vspace*{2mm}

\nin $\bullet$ When $r=1$ ($h=\infty$), we have $M^{(0)}(r=1)=-1$. Again, this is as expected, since the presence of the term $h \sigma_1^z$ in $H'$ will force the magnetisation at site 1 to $-1$ as $h\ra \infty$. This is also the value of the boundary magnetisation 
$M^{(0)}_{bound}(r)$ as $h\ra \infty$. 

\vspace*{2mm}

\begin{figure}[htbp]
\centering
\includegraphics[width=9cm]{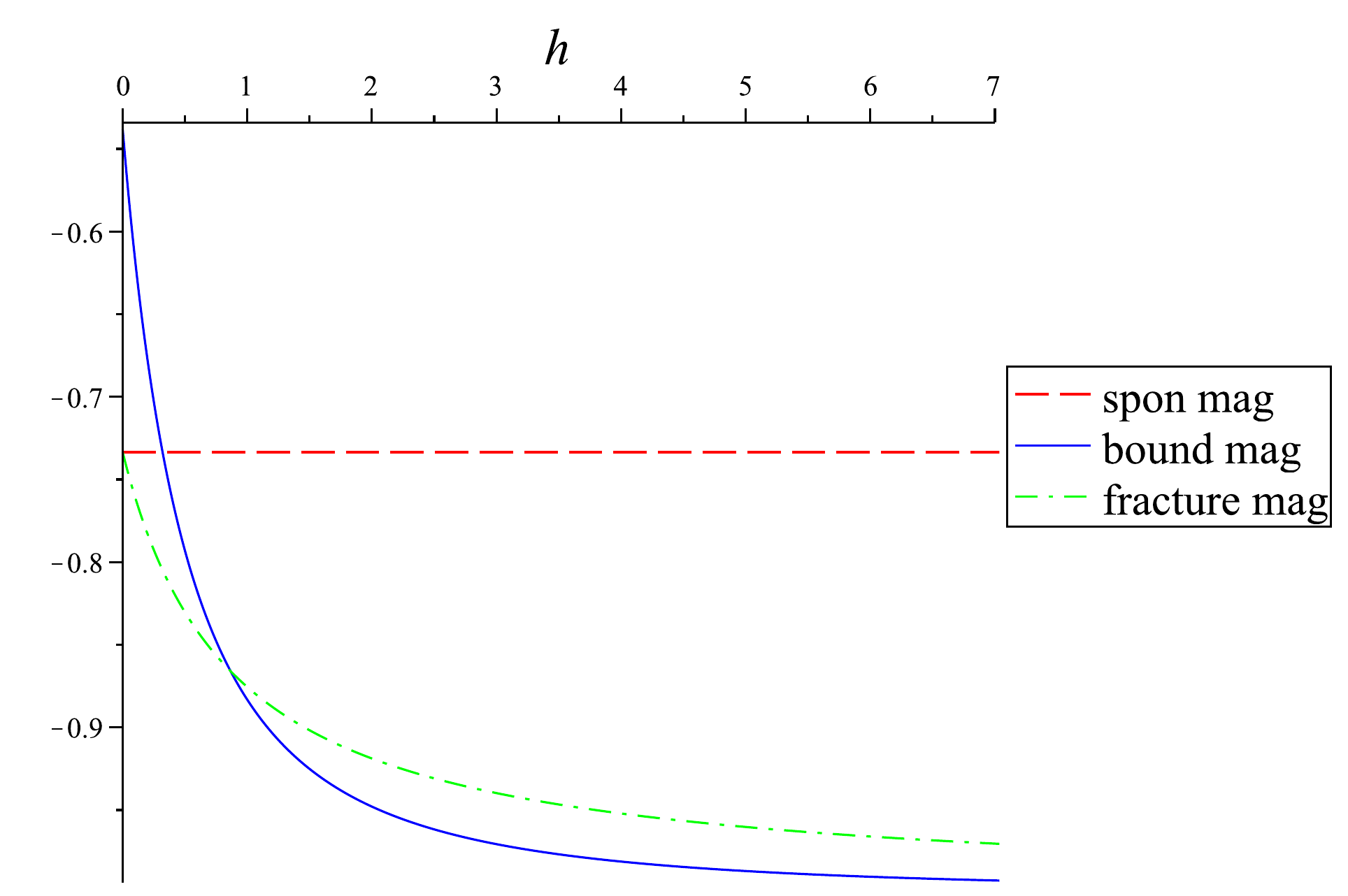}
\caption{The Spontaneous Magnetisation (with added - sign), and the Fracture Magnetisation $M^{(0)}(r)$ and Boundary Magnetisation $M^{(0)}_{bound}(r)$ at site 1 as a Function of Magnetic Field $h$ (at $\Delta=-2$)}
\label{threemags}
\end{figure}

In order to summarise the behaviour, we fix $\Delta=-2$ and plot the two magnetisations $M^{(0)}(r)$ (fracture mag) and $M^{(0)}_{bound}(r)$ (bound mag) as a function of $h(r)$ in 
Figure \ref{threemags}. We also show minus the value of the zero-field bulk spontaneous magnetisation (spon mag) for comparison. The coincidence
of $M^{(0)}(r)$ with the spontaneous magnetisation  at $h=0$, and with $M^{(0)}_{bound}(r)$ at $h_{inv}$ and $h\ra \infty$ is evident. 

\newpage
\section{Summary and Discussion}\setcounter{equation}{0}
In this paper, we have addressed the novel problem of computing XXZ correlation functions in the fractured geometry of Figure \ref{fracorr1}. 
We have expressed these correlation functions in terms of matrix elements $P^{(i)}(\z_1,\z_2,\cdots,\z_N)$ defined  by \mref{cf2} that obey a level 2 boundary qKZ equation. Making use of the free-field realisation, we have derived the general multiple-integral formula \mref{gencfn}. We have specialised this formula to the case relevant to the magnetisation at site 1, and by carrying out a high-order q expansion of the integral have arrived at the conjectural form given by Equation \mref{magform}. This form is similar, but different, to the magnetisation \mref{boundmag} at the corresponding position in the boundary geometry of Figure \ref{boundpfn}. The latter magnetisation was computed in \cite{JKKKM}, in which case the corresponding integral was carried out exactly by summing residues. However, while the final magnetisation expressions are similar, the integrand in our fractured case is considerably more complicated that in the pure boundary case of \cite{JKKKM}. Ultimately, this fact is linked to the additional $(-q)^D$ factor that distinguishes the boundary and fracture correlation functions given by \mref{boundcof} and \mref{fraccof}. The origin of this extra factor lies in the presence of two extra two corner transfer matrices $A^{(i)}_{SE}(\z) A^{(i)}_{SW}(\z)$ in the fractured partition function
 of Figure \ref{fracpfn} when compared to the boundary partition function of  Figure \ref{boundpfn}. This additional factor $(-q)^D$ factor is also responsible for the different level of the boundary qKZ equations in the two cases.

The corner magnetisation in a wedge geometry has been considered before. Conformal field theories in such a geometry were analysed by a conformal mapping from the upper-half plane \cite{0305-4470-17-17-005}. For the Ising model on a triangular lattice, a finite-size scaling approach was taken in \cite{MR775790}, while  the Ising model on a cone was considered in the same paper by taking a trace over a variable number of corner transfer matrices. It would be interesting to try to connect our magnetisation results with these existing calculations in the limit as the wedge angle approaches $2\pi$.

Almost as a bi-product of our analysis we have calculated the overlap  $ {_{(i)}}\!\langle \hb{vac} \vac'_{(i)}$. As we have mentioned in the introduction, our fractured geometry, and in particular this overlap, have recently become objects of interest in the literature on local quantum quenches and their use as a probe of the dynamics of quantum entanglement \cite{CC07,PhysRevLett.106.150404,DubStep11}. So far, most results in this field have been obtained using conformal field theory techniques, and we hope and anticipate that our exact massive lattice model results will be valuable in this arena.

Finally, the vertex operator approach to boundary problems has been generalised in a number of directions  \cite{MR1436155,MR1814559,MR1765615,MR2791123}, and there will be corresponding generalisations of the analysis of fractured models that we have developed here. In particular, the analysis should be generalisable to ABF models \cite{MR1436155} and higher spin models. The former generalisation might be particularly useful in establishing connections with conformal field theory results for wedge geometries.

\subsection*{Acknowledgements}
The author would like to thank Patrick Dorey, Christian Korff, Jorn Mossel, Paul Pearce and Paul Zinn-Justin for useful comments.

\newpage
\begin{appendix}

\section{Bulk and Boundary Weights}\setcounter{equation}{0}
We make use of the following bulk and boundary weights \cite{JKKKM}:
\bea R(\z)=\frac{1}{\kappa(\z)} \bpm 1\\&\frac{(1-\z^2)q}{1-q^2\z^2}& \frac{(1-q^2)\z}{1-q^2\z^2}\\[2mm]
&\frac{(1-q^2)\z}{1-q^2\z^2}&\frac{(1-\z^2)q}{1-q^2\z^2}\\
&&&1
\epm,\quad K(\z;r)=\frac{1}{f(\z;r)}\bpm \frac{1-r\z^2}{\z^2-r}&0\\0&1\epm \lb{rm},\eea 
where \ben
\kappa\z)=\z\frac{(q^4\z^2;q^4)_\infty (q^2\z^{-2};q^4)_\infty }{(q^4\z^{-2};q^4)_\infty(q^2\z^2;q^4)_\infty},\quad
f(\z;r)= \frac{\varphi(\z^{-2};r)}{\varphi(\z^2;r)},\quad \varphi(z;r)= \frac{(q^4 r z;q^4)_\infty(q^6 z^2;q^8)_\infty }
{(q^2 r z;q^4)_\infty (q^8 z^2;q^8)_\infty}.\een
Components are defined by
\ben R(\z) (v_{\ep_1}\ot v_{\ep_2}) = \sli_{\ep'_1,\ep'_2} R^{\ep_1,\ep_2}_{\ep'_1,\ep'_2}(\z) (v_{\ep'_1}\ot v_{\ep'_2}),\quad
K(\z;r) v_{\ep}=\sli_{\ep' }K^{\ep}_{\ep'}(\z;r) v_{\ep'}.\een 
The matrix $R(\z)$ obeys the Yang-Baxter, crossing and unitarity relations, and $K(\z;r)$ obeys the boundary Yang-Baxter, boundary unitarity, and boundary crossing relations shown in \cite{JKKKM}. In this paper, the property that is used most is the crossing symmetry of the R-matrix:
\bea  R^{\ep_1,\ep_2}_{\ep'_1,\ep'_2}(\z) = 
R^{-\ep'_2,\ep_1}_{-\ep_2,\ep'_1}(-q^{-1} \z^{-1})= R^{\ep_2,-\ep'_1}_{\ep'_2,-\ep_1}   (-q^{-1} \z^{-1}).\lb{xing}\eea
We make the further definitions 
\bea K_\bullet(\zeta)=K(\zeta;r),\quad\hb{and} \ws K_\circ(\zeta)=K(-q^{-1}\zeta^{-1};r).\lb{Kdefs}\eea 

\begin{figure}[htbp]\label{btrans}
\centering
\includegraphics[width=16cm]{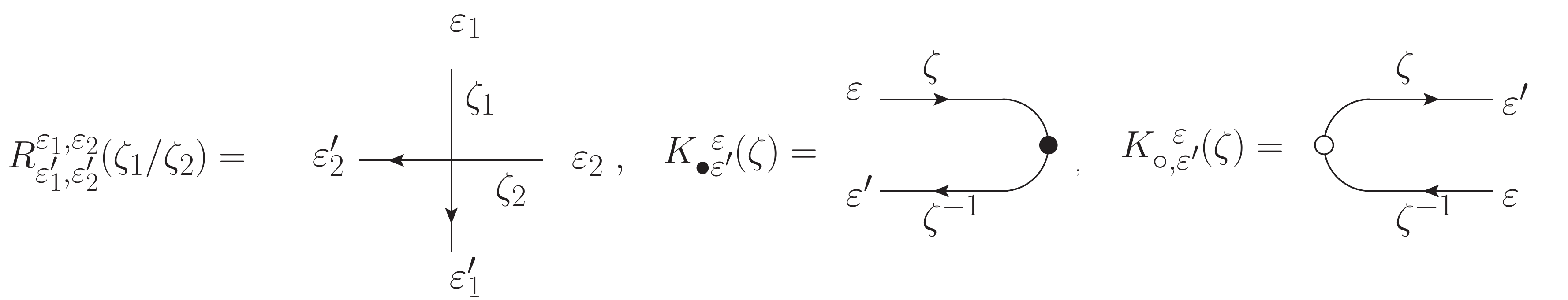}
\caption{Graphical Representation of the Boltzmann Weights}
\label{graphrep}
\end{figure}

The graphical representation of these matrices used throughout the current paper is that shown in Figure \ref{graphrep}.

\newpage
\section{The Integral Formula for Correlation Functions}\setcounter{equation}{0}
In this appendix, we give the details of the free-field calculation of ${_B}\langle i|(-q)^D |i\rangle_B$ and of \bea 
\hspace*{-5mm}P^{(i)}(\zeta_1,\zeta_2,\cdots,\zeta_N)_{\ep_1,\ep_2,\cdots,\ep_m}=\frac{1}{{_B}\langle i | (-q)^{D} |i\rangle_B} {_B}\langle i |(-q)^{D} \Phi_{\ep_1}(\zeta_1)\Phi_{\ep_2}(\zeta_2) \cdots \Phi_{\ep_N}(\zeta_N)|i\rangle_B,\lb{cfn10}\eea
when $N$ is even. We closely follow the approach and notational conventions of \cite{JKKKM} in which ${_B}\langle i |i\rangle_B$ and 
$\frac{1}{{_B}\langle i |i\rangle_B} {_B}\langle i | \Phi_{\ep_1}(\zeta_1)\Phi_{\ep_2}(\zeta_2) \cdots \Phi_{\ep_N}(\zeta_N)|i\rangle_B$ are calculated; the key difference is that we replace ${_B}\langle i|$ of \cite{JKKKM} with ${_B}\langle i |(-q)^D$.

The states $\ket{i}_B$ and ${_B}\bra{i}$ are expressed in the free-field realisation of \cite{JKKKM} as 
\ben &&|i\rangle_B = e^{F_i} \ket{\gL_i},\quad {_B}\langle i|=\bra{\gL_i} e^{G_i},\\
&&\hb{where}\ws  F_i =\frac{1}{2}\sli_{n=1}^\infty \frac{n \alpha_n}{[2n][n]}a_{-n}^2+ \sli_{n=1}^\infty \beta_n^{(i)} a_{-n},\quad
G_i =\frac{1}{2}\sli_{n=1}^\infty \frac{n \gamma_n}{[2n][n]}a_{n}^2+ \sli_{n=1}^\infty \delta_n^{(i)} a_{n} .\een
Here, the coefficients are given by
\ben 
\alpha_n&=&-q^{6n},\ws \gamma_n=-q^{-2n},\\
\beta_n^{(0)}&=& -\theta_n \frac{q^{5n/2}(1-q^n)}{[2n]}-\frac{q^{7n/2} r^n}{[2n]} ,\ws
\delta_n^{(0)}= \theta_n \frac{q^{-3n/2}(1-q^n)}{[2n]}-\frac{q^{-5n/2} r^n}{[2n]},\\
\beta_n^{(1)}&=& -\theta_n 
\frac{q^{5n/2}(1-q^n)}{[2n]}
+\frac{q^{3n/2} r^{-n}}{[2n]} ,\ws
\delta_n^{(1)}=\theta_n \frac{q^{-3n/2}(1-q^n)}{[2n]}+ 
\frac{q^{-n/2} r^{-n}}{[2n]},\een
where \ben \\[-12mm]
[a]:=\frac{q^a-q^{-a}}{q-q^{-1}},\quad \hb{and}\quad \theta_n:=\begin{cases} 
1& \hb{if  }n\hb{  is even,}\\ 
0& \hb{if  }n\hb{  is odd.}\end{cases}\een

Given the free-field realisation properties $(-q)^{-D} a_{n} (-q)^{D} = q^{2n} a_n$ and $\bra{\gL_i}(-q)^D=\bra{\gL_i}$, it follows that we have
\ben {_B}\langle i |(-q)^D&=& \bra{\gL_i} e^{G'_i},\quad
\hb{where}\ws G'_i=\sli_{n=1}^\infty \frac{n \gamma'_n}{[2n][n]}a_{n}^2+ \sli_{n=1}^\infty \delta_n^{'(i)} a_{n},\quad
\gamma'_n= q^{4n} \gamma_n,\ws \delta^{'(i)}_n=q^{2n} \delta^{(i)}_n.\een
Now we closely follow the procedure of Section 4 and Appendix C of \cite{JKKKM} with the change of $\gamma_n\ra \gamma'_n$ and $\delta_n^{(i)} \ra \delta_n^{'(i)}$. 
In this way, we obtain 
\bea
 _B \bra{i}(-q)^D |i\rangle_B\hspace*{-2mm}&=&\hspace*{-2mm} \prod_{m=1}^\infty \frac{1}{(1-\alpha_m \gamma'_m)^{\half}}\, \exp \left(\half \sli_{n=1}^\infty 
\frac{[2n][n]}{n(1-\alpha_n\gamma'_n)}(\,\gamma'_n (\beta_n^{(i)})^2+2\delta_n^{'(i)} 
\beta_n^{(i)} + (\delta_n^{'(i)})^2\alpha_n) \right).\nn\\ \lb{norm}\\[-12mm]\nn\eea
The sums here  can be written as infinite products using the identity
\ben \exp\left( -\sli_{n=1}^\infty \frac{z^{n}}{n (1-q^{m_1 n})(1-q^{m_2 n}) \cdots  (1-q^{m_N n})}\right)=(z;q^{m_1},q^{m_2},\cdots,q^{m_N})_\infty,\quad |z|<1.\een
Carrying this procedure out and making use of standard q-product identities, as well as the less-standard, but easily-derived, identity
\ben (a;b,b)_\infty  =  (ab;b,b^2)_\infty^2 \,(a;b^2)_\infty,\een
we find \ben _B \bra{0}(-q)^D |0\rangle_B&=&\frac{(r^2 q^{10};q^8,q^8)^2_\infty}{ (r^2 q^{12};q^8,q^8)^2_\infty} 
\frac{(r^2 q^{6};q^4,q^8)_\infty}{(r^2 q^{4};q^4,q^8)_\infty} 
\frac{( q^{14} ;q^8,q^8)_\infty}{( q^{10};q^8,q^8)_\infty},\een
and that $_B \bra{1}(-q)^D |1\rangle_B$ is given by the same formula with the substitution $r\ra r^{-1}$.
Using the expressions \ben  _B \bra{0}0\rangle_B= \frac{(q^4 r^2;q^8)_\infty}{(q^2 r^2;q^8)_\infty (q^6;q^8)_\infty},
\quad  _B \bra{1}1\rangle_B= \frac{(q^4 r^{-2};q^8)_\infty}{(q^2 r^{-2};q^8)_\infty (q^6;q^8)_\infty}\een of \cite{JKKKM},
and the definition \mref{eqn:chi} of $\chi$, we obtain the result for ${_{(i)}}\!\langle \hb{vac} \vac'_{(i)}$ given by Equation \mref{ovlap}.

Let us carry on to calculate the correlation function \mref{cfn10}.
Defining the set $A$ by \ben A=\{j|1\leq j\leq N, \ep_j=+1\},\een we again follow the method of Section 4 of \cite{JKKKM} to obtain\footnote{There is a minor typographical error in (4.4) of \cite{JKKKM} in which an additional factor is present on the left-hand side. However,  
this error is not carried through and the final expression (4.8) in \cite{JKKKM} is correct.}
\ben P^{(i)}(\z_1,\z_2,\cdots,\z_N)_{\ep_1,\ep_2,\cdots,\ep_N}&=&(-q^3)^{N^2/4+iN/2-\sli_{a\in A } a}(1-q^2)^{N/2} \pl_{j=1}^N \z_j^{\frac{1+\ep_j}{2}-j+N+i} \pl_{j<k} 
\frac{(q^2z_k/z_j;q^4)_\infty}{(q^4 z_k/z_j;q^4)_\infty}\\
&&\hspace*{-15mm}\times \pl_{a\in A} \oint_{C_a} \frac{dw_a}{2\pi\sqrt{-1}} w_a^{1-i} \frac{\pl_{a<b} (w_a-w_b)(w_a-q^2 w_b) }
{\pl_{j\leq a} (z_j-q^{-2} w_a) \pl_{a\leq j} (w_a-q^4 z_j)  } I^{'(i)}(\{z_j\},\{w_a\}),\een
where the contour $C_a$ is given by $q^4 z_j<w_a<q^2z_j$ (recall - the notation $A<w_a<B$ means that the pole at $A$ lies inside, and pole at $B$ lies outside the anticlockwise $w_a$ contour) and 
\bea   I^{'(i)}(\{z_j\},\{w_a\})&=&\exp\left(\sli_{n=1}^\infty \frac{[2n][n]}{n} \frac{1}{1-\alpha_n\gamma'_n} \left\{ \half\gamma_n' X_n^2-\alpha_n \gamma'_n X_n Y_n \right.\right. \nn\\&& \left.\left.+\half \alpha_n Y_n^2 +(\delta^{'(i)}_n +\gamma'_n \beta_n^{(i)}) X_n - (\beta_n^{(i)}+\alpha_n \delta^{'(i)}_n) Y_n \right\} \right),\lb{cfn7}\eea with
\ben 
X_n&=&\frac{q^{7n/2}}{[2n]} \sli_{j=1}^N z_j^n - \frac{q^{n/2}}{[n]} \sli_{a}w_a^n,\quad Y_n=\frac{q^{-5n/2}}{[2n]} \sli_{j=1}^N z_j^{-n} - 
\frac{q^{n/2}}{[n]} \sli_{a}w_a^{-n}.
\een
Computing the sums as before we find 
\bea &&\hspace*{-5mm}I^{'(i)}(\{z_j\},\{w_a\})=F^{'(i)}(\{z_j\},\{w_a\})\nn \\  &&\hspace*{-5mm}\times \prod_{j<k}\frac{ (q^{10}z_j z_k;q^4,q^8)_\infty }{(q^{12} z_j z_k;q^4,q^8)_\infty}
  \frac{ (q^{2}/(z_j z_k);q^4,q^8)_\infty }{(q^{4}/( z_j z_k);q^4,q^8)_\infty}
\prod_{j,k} \frac{ (q^{10}z_j/ z_k;q^4,q^8)_\infty }{(q^{12} z_j /z_k;q^4,q^8)_\infty}
\prod_j\frac{(q^{14} z_j^2;q^8,q^{8})_\infty}{(q^{16} z_j^2;q^8,q^{8})_\infty} 
\frac{(q^{6}/ z_j^2;q^8,q^{8})_\infty}{(q^{8} /z_j^2;q^8,q^{8})_\infty} \nn \\&&\hspace*{-5mm}\times
\prod_{j,a} \frac{1}{(q^6z_j w_a;q^8)_\infty (q^4/(z_j w_a);q^8)_\infty (q^{12} z_j/ w_a;q^8)_\infty (q^{6} 
w_a/z_j;q^8)_\infty }
\prod_a (q^2 w_a^2;q^8)_\infty(q^6/ w_a^2;q^8)_\infty \lb{cfn8}\\ 
&&\hspace*{-5mm}\times \prod_{a<b}
(q^2w_a w_b;q^8)_\infty (q^4 w_a w_b;q^8)_\infty (q^6/(w_a w_b);q^8)_\infty (q^8/( w_a w_b);q^8)_\infty 
\prod_{a,b} (q^8 w_a /w_b;q^8)_\infty(q^{10} w_a /w_b;q^8)_\infty,\nn\eea
where 
\ben 
&&F^{'(0)}(\{z_j\},\{w_a\})=\\ &&\prod_{j} \frac{ (q^{4} rz_j;q^4,q^8)_\infty (q^{12} rz_j;q^4,q^8)_\infty   }{(q^6 r z_j;q^4,q^4)_\infty}
 \frac{ (q^{4} r/z_j;q^4,q^8)^2_\infty   }{(q^2 r /z_j;q^4,q^4)_\infty}
\prod_a \frac{(q^6 rw_a;q^8)_\infty (q^4 r/w_a;q^8)_\infty }{(rw_a;q^8)_\infty (q^6 r/w_a;q^8)_\infty},\\
&&F^{'(1)}(\{z_j\},\{w_a\})=\\ &&\prod_{j} 
 \frac{ (1/(r z_j);q^4,q^8)_\infty (q^{8} /(r z_j);q^4,q^8)_\infty   }{(q^2/(rz_j);q^4,q^4)_\infty}
\frac{ (q^{8} z_j/r;q^4,q^8)^2_\infty   }{(q^6  z_j/r;q^4,q^4)_\infty}
\prod_a \frac{(q^2 w_a/r;q^8)_\infty (q^8 /(rw_a);q^8)_\infty }{(q^4 w_a/r;q^8)_\infty (q^2 /(rw_a);q^8)_\infty}.
\een
The full contour $C_a$ is now fixed by the previous requirement $q^4 z_j <w_a<q^2 z_j$ \,supplemented by the requirement that the contour be consistent with the region of validity of each of the sums contributing to the last expression for $I^{'(i)}(\{z_j\},\{w_a\})$. Hence, if a term $(w_a/A;q^8)_\infty$ appears in the denominator of $I^{'(i)}(\{z_j\},\{w_a\})$, we insist that
$ w_a< A q^{-8n}$ for $n\geq 0$; if a term $(B/w_a;q^8)_\infty$ appears we insist that $Bq^{8n}<w_a$ for $n\geq 0$.

\end{appendix}
\newpage
\baselineskip=12pt

\begin{thebibliography}{10}

\bibitem{Bethe31}
H.~Bethe.
\newblock {On the Theory of Metals, I. Eigenvalues and Eigenfunctions of a
  Linear Chain of Atoms}.
\newblock {\em Z. Physik}, 71:205--26, 1931.
\newblock English translation appears in {\it Selected Works of Hans A. Bethe
  With Commentary}, H. Bethe (World Scientific, Singapore, 1996).

\bibitem{MR2583103}
B.~M. McCoy.
\newblock {\em {Advanced Statistical Mechanics}}, volume 146 of {\em
  International Series of Monographs on Physics}.
\newblock Oxford University Press, Oxford, 2010.

\bibitem{Korepin}
V.E.Korepin, N.M.Bogoliubov, and A.G.Izergin.
\newblock {\em Quantum Inverse Scattering Method and Correlation Functions}.
\newblock CUP, Cambridge, 1993.

\bibitem{Bax82}
R.~J. Baxter.
\newblock {\em Exactly Solved Models in Statistical Mechanics}.
\newblock Academic, London, 1982.

\bibitem{Aff89}
I.~Affleck.
\newblock {\em Field Theory Methods and Quantum Critical Phenomena}.
\newblock North Holland, Amsterdam, 1990.
\newblock Les Houches, Session XLIX, 1988, {\it Champs, Cordes et
  Ph\'enom\`enes Critiques}, Ed. E. Br\'ezin and J. Zinn-Justin.

\bibitem{Jim85}
M.~Jimbo.
\newblock {A q-analogue of U(gl(N+1)), Hecke algebra, and the Yang-Baxter
  equation }.
\newblock {\em Lett. Math. Phys.}, 10:63, 1985.

\bibitem{Dri85}
V.~G. Drinfeld.
\newblock {Hopf algebras and the quantum Yang-Baxter equation}.
\newblock {\em Soviet Math. Doklady}, 32:254, 1985.

\bibitem{Daval}
B.~Davies, O.~Foda, M.~Jimbo, T.~Miwa, and A.~Nakayashiki.
\newblock {Diagonalization of the XXZ Hamiltonian by Vertex Operators}.
\newblock {\em Comm. Math. Phys.}, 151:89--153, 1993.

\bibitem{JM}
M.~Jimbo and T.~Miwa.
\newblock {\em Algebraic Analysis of Solvable Lattice Models}.
\newblock CBMS Regional Conference Series in Mathematics, vol. 85. Amer. Math.
  Soc., 1994.

\bibitem{Goff95}
J~P Goff, D~A Tennant, and S~E Nagler.
\newblock {Exchange mixing and soliton dynamics in the quantum spin chain
  CsCoCl$_3$}.
\newblock {\em Phys Rev}, B 52:15992, 1995.

\bibitem{Nagler83}
S.~E. Nagler, W.J.L. Buyers, R.L. Armstrong, and B.~Briat.
\newblock {S}olitons in the one-dimensional anti-ferromagnet {C}s{C}o{B}r$_3$.
\newblock {\em Phys. Rev.}, B 28:3873--3885, 1983.

\bibitem{PhysRevLett.91.090402}
L.-M. Duan, E.~Demler, and M.~D. Lukin.
\newblock Controlling spin exchange interactions of ultracold atoms in optical
  lattices.
\newblock {\em Phys. Rev. Lett.}, 91(9):090402, 2003.

\bibitem{Beisertreview}
N.~Beisert et~al.
\newblock {Review of AdS/CFT Integrability: An Overview}.
\newblock arXiv:1012.3982v4, 2010.

\bibitem{MR1702631}
A.~G. Izergin, N.~Kitanine, J.~M. Maillet, and V.~Terras.
\newblock Spontaneous magnetization of the {$XXZ$} {H}eisenberg spin-{$\frac
  12$} chain.
\newblock {\em Nuclear Phys. B}, 554(3):679--696, 1999.

\bibitem{MR1741654}
N.~Kitanine, J.~M. Maillet, and V.~Terras.
\newblock Correlation functions of the {$XXZ$} {H}eisenberg spin-{${1\over2}$}
  chain in a magnetic field.
\newblock {\em Nuclear Phys. B}, 567(3):554--582, 2000.

\bibitem{JKKMW}
M.~Jimbo, R.~Kedem, H.~Konno, T.~Miwa, and R.A. Weston.
\newblock Difference equations in spin chains with a boundary.
\newblock {\em Nucl. Phys.}, B448:429--456, 1995.

\bibitem{1742-5468-2007-06-P06005}
V.~Eisler and I.~Peschel.
\newblock Evolution of entanglement after a local quench.
\newblock {\em JSTAT}, (06):P06005, 2007.

\bibitem{CC07}
P.~Calabrese and J.~Cardy.
\newblock {Entanglement and Correlation functions following a local quench: a
  conformal field theory approach}.
\newblock {\em JSTAT}, (10):P10004, 2007.

\bibitem{1742-5468-2011-03-L03002}
J.~Dubail and J-M. St\'ephan.
\newblock Universal behavior of a bipartite fidelity at quantum criticality.
\newblock {\em JSTAT}, (03):L03002, 2011.

\bibitem{DubStep11}
J-M. St\'ephan and J.~Dubail.
\newblock {Local quantum quenches in critical one-dimensional systems:
  entanglement, Loschmidt echo, and light-cone effects}, 2011.
\newblock arXiv:1105.4846.

\bibitem{PhysRevLett.106.150404}
J.~L. Cardy.
\newblock {Measuring Entanglement Using Quantum Quenches}.
\newblock {\em Phys. Rev. Lett.}, 106:150404, 2011.

\bibitem{0305-4470-16-15-026}
J.~L. Cardy.
\newblock {Critical Behaviour at an Edge}.
\newblock {\em Journal of Physics A: Mathematical and General}, 16(15):3617,
  1983.

\bibitem{0305-4470-17-17-005}
J.~L. Cardy and S.~Redner.
\newblock {Conformal invariance and self-avoiding walks in restricted
  geometries}.
\newblock {\em Journal of Physics A: Mathematical and General}, 17(17):L933,
  1984.

\bibitem{MR775790}
M.~N. Barber, I.~Peschel, and P.~A. Pearce.
\newblock Magnetization at corners in two-dimensional {I}sing models.
\newblock {\em J. Statist. Phys.}, 37(5-6):497--527, 1984.

\bibitem{JKKKM}
M.~Jimbo, R.~Kedem, T.~Kojima, H.~Konno, and T.~Miwa.
\newblock {XXZ chain with a boundary}.
\newblock {\em Nucl. Phys.}, B441 [FS]:437--470, 1995.

\bibitem{SKl87}
E.~K. Sklyanin.
\newblock Boundary conditions for integrable quantum systems.
\newblock {\em J. Phys. A}, 21:2375--2389, 1988.

\bibitem{Cher}
I.~V. Cherednik.
\newblock Factorizing particles on a half-line and root systems.
\newblock {\em Theor. Math. Phys.}, 61:977--983, 1984.

\bibitem{MR2186202}
P.~Di~Francesco.
\newblock Boundary q{KZ} equation and generalized {R}azumov-{S}troganov sum
  rules for open {IRF} models.
\newblock {\em JSTAT}, (11):P11003, 2005.

\bibitem{MR2367185}
P.~Di~Francesco and P.~Zinn-Justin.
\newblock Quantum {K}nizhnik-{Z}amolodchikov equation: reflecting boundary
  conditions and combinatorics.
\newblock {\em JSTAT}, (12):P12009, 2007.

\bibitem{FR}
I.~B. Frenkel and N.~Yu Reshetikhin.
\newblock {Quantum Affine Algebras and Holonomic Difference Equations}.
\newblock {\em Comm. Math. Phys.}, 146:1--60, 1992.

\bibitem{MR926391}
F.~C. Alcaraz, M.~N. Barber, M.~T. Batchelor, R.~J. Baxter, and G.~R.~W.
  Quispel.
\newblock Surface exponents of the quantum {$XXZ$}, {A}shkin-{T}eller and
  {P}otts models.
\newblock {\em J. Phys. A}, 20(18):6397--6409, 1987.

\bibitem{MR1043392}
V.~Pasquier and H.~Saleur.
\newblock Common structures between finite systems and conformal field theories
  through quantum groups.
\newblock {\em Nuclear Phys. B}, 330(2-3):523--556, 1990.

\bibitem{NIST}
F.~W.~J. Olver, D.~W. Lozier, R.~F. Boisvert, and C.~W. Clark, editors.
\newblock {\em {NIST Handbook of Mathematical Functions}}.
\newblock U.S. Department of Commerce National Institute of Standards and
  Technology, Washington, DC, 2010.

\bibitem{MR1436155}
T.~Miwa and R.~Weston.
\newblock Boundary {ABF} models.
\newblock {\em Nuclear Phys. B}, 486(3):517--545, 1997.

\bibitem{MR1814559}
T.~Kojima.
\newblock The massless {$XXZ$} chain with a boundary.
\newblock {\em Internat. J. Modern Phys. A}, 16(3):409--424, 2001.

\bibitem{MR1765615}
H.~Furutsu and T.~Kojima.
\newblock The {$U_q(\widehat{\rm sl}_n)$} analogue of the {$XXZ$} chain with a
  boundary.
\newblock {\em J. Math. Phys.}, 41(7):4413--4436, 2000.

\bibitem{MR2791123}
T.~Kojima.
\newblock Diagonalization of infinite transfer matrix of boundary
  {$U_{q,p}(A^{(1)}_{N-1})$} face model.
\newblock {\em J. Math. Phys.}, 52(1):013501, 26, 2011.

\end{thebibliography}

\end{document}